\documentclass[sigconf]{acmart}

\AtBeginDocument{%
  }

\usepackage{tcolorbox}

\usepackage{bbding}
\usepackage{subfigure}
\usepackage[boxed,ruled,lined]{algorithm2e}
\usepackage{multirow}
\usepackage{makecell}

\newcommand{\argmin}{\operatornamewithlimits{arg\,min}}
\usepackage{arydshln} 
\usepackage{enumerate}
\usepackage{enumitem}
\usepackage{makecell}

\newcommand{\recitem}{$\langle \text{R}_{\text{I}} \rangle$\xspace}
\newcommand{\srcquery}{$\langle \text{S}_{\text{Q}} \rangle$\xspace}
\newcommand{\srcitem}{$\langle \text{S}_{\text{I}} \rangle$\xspace}

\newcommand{\sharecodea}{$\text{M}_1$}
\newcommand{\sharecodeb}{$\text{M}_2$}
\newcommand{\reccodea}{$\text{R}_1$}
\newcommand{\reccodeb}{$\text{R}_2$}
\newcommand{\srccodea}{$\text{S}_1$}
\newcommand{\srccodeb}{$\text{S}_2$}

\newcommand{\srcandrec}{S\&R\xspace}

\newcommand{\ourname}{GenSAR\xspace}

\newcommand{\paratitle}[1]{\vspace{0.8ex}\noindent\textbf{#1}}

\setcopyright{acmlicensed}
\copyrightyear{2018}
\acmYear{2018}
\acmDOI{XXXXXXX.XXXXXXX}
\acmConference[Conference acronym 'XX]{Make sure to enter the correct
  conference title from your rights confirmation emai}{June 03--05,
  2018}{Woodstock, NY}
\acmISBN{978-1-4503-XXXX-X/18/06}

\begin{document}

\title{Unified Generative Search and Recommendation}

\author{Teng Shi}
\affiliation{%
\institution{Renmin University of China}
  \city{Beijing}\country{China}
  }
\email{shiteng@ruc.edu.cn}

\author{Jun Xu}
\authornote{Corresponding authors. Work partially done at Engineering Research Center of Next-Generation Intelligent Search and Recommendation, Ministry of Education. \\
Work done when Teng Shi was the intern at Kuaishou.
}
\author{Xiao Zhang}
\affiliation{
  \institution{Renmin University of China}
  \city{Beijing}\country{China}
  }
\email{{junxu,zhangx89}@ruc.edu.cn}

\author{Xiaoxue Zang}
\author{Kai Zheng}
\affiliation{
  \institution{Kuaishou Technology Co., Ltd.}
  \city{Beijing}\country{China}
  }
\email{xxic666@126.com}
\email{zhengk92@gmail.com}

\author{Yang Song}
\affiliation{
  \institution{Kuaishou Technology Co., Ltd.}
  \city{Beijing}\country{China}
}
\email{ys@sonyis.me}

\author{Enyun Yu}
\affiliation{
    \institution{Independent}
  \city{Beijing}\country{China}
}
\email{yuenyun@126.com}

\renewcommand{\shortauthors}{Teng Shi et al.}

\begin{abstract}
Modern commercial platforms typically offer both search and recommendation functionalities to serve diverse user needs, making joint modeling of these tasks an appealing direction. While prior work has shown that integrating search and recommendation can be mutually beneficial, it also reveals a performance trade-off: enhancements in one task often come at the expense of the other. This challenge arises from their distinct information requirements: search emphasizes semantic relevance between queries and items, whereas recommendation depends more on collaborative signals among users and items. Effectively addressing this trade-off requires tackling two key problems: (1) integrating both semantic and collaborative signals into item representations, and (2) guiding the model to distinguish and adapt to the unique demands of search and recommendation. The emergence of generative retrieval with Large Language Models (LLMs) presents new possibilities. This paradigm encodes items as identifiers and frames both search and recommendation as sequential generation tasks, offering the flexibility to leverage multiple identifiers and task-specific prompts. In light of this, we introduce GenSAR, a unified generative framework for balanced search and recommendation. Our approach designs dual-purpose identifiers and tailored training strategies to incorporate complementary signals and align with task-specific objectives. Experiments on both public and commercial datasets demonstrate that GenSAR effectively reduces the trade-off and achieves state-of-the-art performance on both tasks.

\end{abstract}



\maketitle

\begin{figure}[t]
     \centering
     \subfigure[Trade-off between \srcandrec]{
        \label{fig:intro_trade_off}
        \includegraphics[width=0.45\columnwidth]{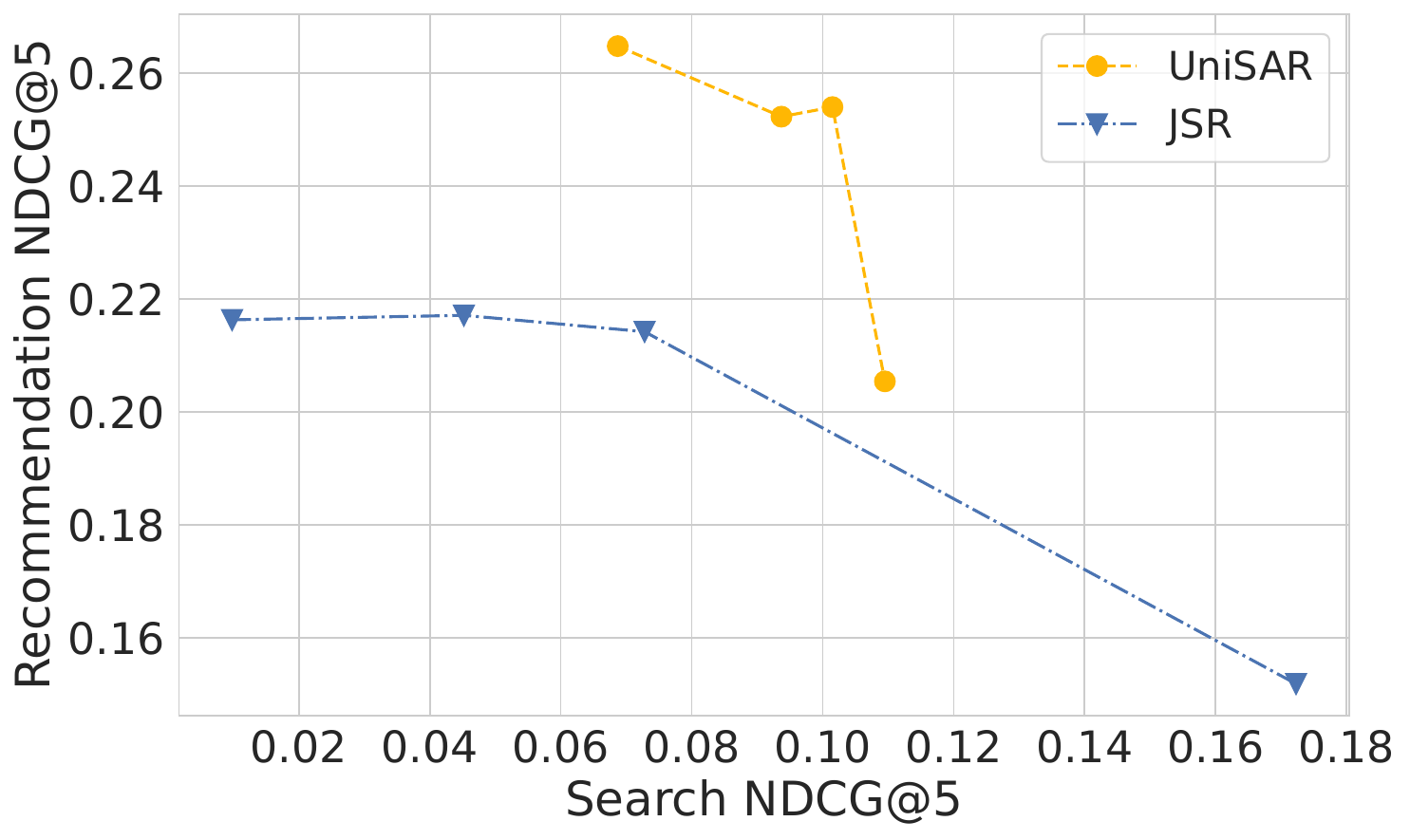}
     }
    \subfigure[Performance of different embeddings]{
        \label{fig:intro_emb}
        \includegraphics[width=0.45\columnwidth]{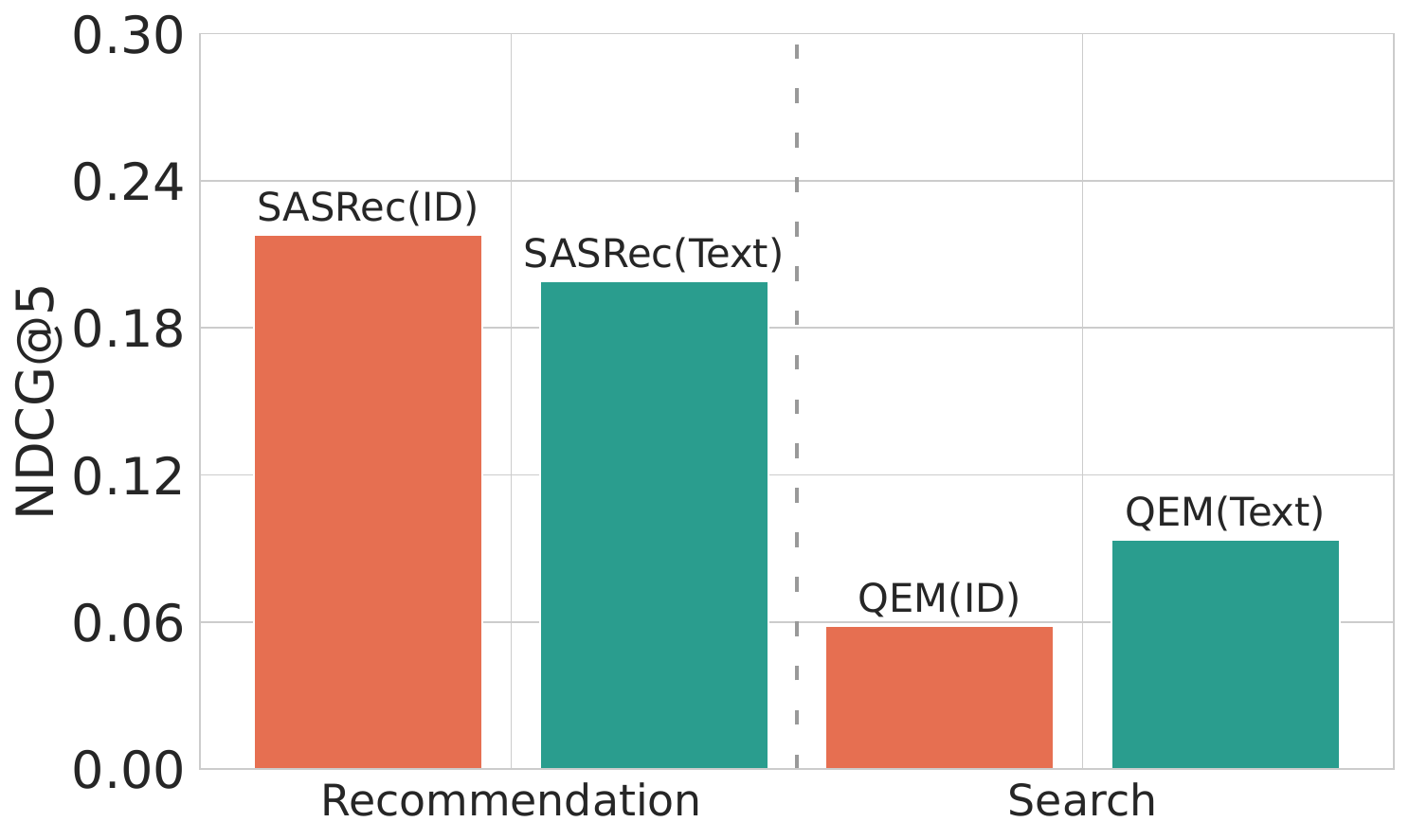}
     }
     \vspace{-12px}
     \caption{
     Empirical analysis on the Commercial dataset: (a) A trade-off between \srcandrec is observed in representative joint \srcandrec methods, JSR~\cite{JSR} and UniSAR~\cite{UniSAR}. (b) The performance of the sequential recommendation model SASRec~\cite{SASREC} and the product search model QEM~\cite{ai2019zero}, using ID and text embeddings, respectively. 
     }
     \label{fig:intro_example}
     \vspace{-0.5cm}
\end{figure}

\section{Introduction}

To facilitate the diverse ways of information access, 
many commercial platforms, such as e-commerce, video, and music platforms, offer both search~\cite{ai2017learning,ai2019zero,bi2020transformer,CoPPS} and recommendation~\cite{zhang2024saqrec,zhang2024qagcf,zhang2024modeling,zhang2024model,zhang2025testtimealignmenttrackinguser,tang2025thinkrecommendunleashinglatent} (\textbf{\srcandrec}) services. This provides an opportunity for joint modeling of \srcandrec, enabling better user interest modeling and enhancing the performance of both tasks.

Many studies have explored joint modeling of \srcandrec, including: leveraging recommendation to enhance search~\cite{ai2017learning,ai2019zero,bi2020transformer,CoPPS}, using search to enhance recommendation~\cite{Query_SeqRec,IV4REC,SESRec,wang2024enhancing}, and unified \srcandrec modeling~\cite{JSR,JSR2,USER,xie2024unifiedssr,UniSAR}. Although these studies have demonstrated that \srcandrec can mutually enhance each other, they have also identified a trade-off when the model serves both tasks simultaneously~\cite{UniSAR}. Specifically, when the recommendation performance improves, the search performance tends to degrade, and vice versa. 
Empirical analysis of the representative methods of JSR~\cite{JSR} and UniSAR~\cite{UniSAR} based on a \srcandrec dataset collected from a real commercial platform also confirmed the performance trade-off, as shown in Figure~\ref{fig:intro_trade_off}. 
More details please refer to Section~\ref{sec:exp:dataset}.

Analysis also showed that the trade-off is rooted in the different information requirements of \srcandrec.
Search typically focuses more on the semantic relevance between queries and items, with traditional search models often based on pre-trained language models~\cite{xiong2020approximate,izacard2021unsupervised,xiao2024c}. In contrast, recommendation heavily relies on collaborative information, where ID-based recommendation can yield excellent results~\cite{yuan2023go,he2017neural,SASREC}. 
Figure~\ref{fig:intro_emb} shows an empirical validation where the \srcandrec performances with ID- and Text-only embeddings are shown. The ID embeddings are randomly initialized and trained, containing collaborative information, while the Text embeddings are trained with BGE~\cite{xiao2024c} and then reduced to the same dimensionality as that of the ID embeddings, containing semantic information. From Figure~\ref{fig:intro_emb}, we found that recommendation relies more on collaborative information while search focuses more on semantic~information.

Therefore, balancing the semantic information required for search and the collaborative information needed for recommendation becomes a key issue in joint \srcandrec modeling. It is non-trivial and faces two major challenges:
(1) How to incorporate both semantic and collaborative information in item representations. Existing joint \srcandrec models typically assign a single representation to each item, making it difficult to capture both types of information effectively; (2) How to let the model understand the difference 
in information requirements
of \srcandrec during training. Current joint models often treat \srcandrec tasks identically, without differentiating them during training. This makes it challenging for the model to grasp their distinct requirements.

Recently, Large Language Model~(LLM)~\cite{zhao2023survey}-based generative retrieval for search~\cite{tay2022transformer,zhuang2022bridging} and recommendation~\cite{geng2022recommendation,TIGER,LC_Rec} have garnered significant attention.
This provides a solution to the aforementioned challenges: (1) Generative retrieval assigns an identifier (a sequence of tokens) to each item, allowing us to assign multiple identifiers to each item to balance semantic and collaborative information; (2) Generative retrieval formulates both \srcandrec as sequence-to-sequence (Seq2Seq) tasks, enabling the unification of different \srcandrec tasks and helping the model better understand the distinct requirements of each task.

Based on this, we propose \textbf{\ourname}, which unifies balanced search and recommendation through generative retrieval, thereby alleviating the trade-off between \srcandrec to better enhance each other.
Firstly, we design a joint \srcandrec identifier that integrates both semantic and collaborative information. Building on the RQ-VAE~\cite{TIGER,LC_Rec} method, we employ shared codebooks for both semantic and collaborative information, alongside specific codebooks for each. 
As a result, items from search are represented by semantic codes, while items from recommendation are represented by collaborative codes. These two codes share a common portion to capture shared information while also retaining distinct parts to preserve the unique characteristics of semantic and collaborative information.
Secondly, we design the joint \srcandrec training tasks. We prepend a token representing the behavior type to the item identifier and then input the user's \srcandrec history into the LLM (with the user query also provided for search). 
Different prompts are used to guide LLMs to predict the next recommended item, the next searched query, and the next searched item, enabling the model to understand the distinct requirements for \srcandrec.

The major contributions of the paper are summarized as follows:

\noindent\textbf{$\bullet $}~We verified the existence of the trade-off between \srcandrec, and identified that this trade-off arises from the different information requirements of \srcandrec. Additionally, we have analyzed the challenges in balancing semantic and collaborative information needed for~\srcandrec.

\noindent\textbf{$\bullet $}~We propose \ourname, which unifies balanced \srcandrec through generative retrieval. We designed a joint \srcandrec identifier to balance semantic and collaborative information, and developed joint training tasks to help the model understand the different requirements of each task.    

\noindent\textbf{$\bullet $}~Experimental results on two datasets validate the effectiveness of \ourname. \ourname not only surpasses traditional \srcandrec models but also outperforms generative \srcandrec models.

\section{Related Work}

\paratitle{Joint Search and Recommendation.}
Joint modeling of \srcandrec has attracted increasing attention in recent years and can be broadly categorized into three types:
(1) Enhancing search with recommendation~\cite{ai2017learning,ai2019zero,bi2020transformer,CoPPS}, such as TEM~\cite{bi2020transformer}, which uses Transformers to model user preferences, and CoPPS~\cite{CoPPS}, which applies contrastive learning to address data sparsity.
(2) Enhancing recommendation with search~\cite{Query_SeqRec,IV4REC,SESRec,wang2024enhancing}, e.g., SESRec~\cite{SESRec}, which disentangles similar and dissimilar interests from both histories.
(3) Unified modeling of \srcandrec~\cite{JSR,JSR2,USER,SRJgraph,xie2024unifiedssr,zhang2024unified,UniSAR}, such as JSR~\cite{JSR,JSR2} with joint loss and UniSAR~\cite{UniSAR}, which models behavior transitions.
While these works show mutual benefits between \srcandrec, they also reveal a trade-off~\cite{UniSAR,shen2024survey}. This paper addresses that trade-off within a generative retrieval framework.

\paratitle{Generative Search and Recommendation.}
With the rise of Large Language Models (LLMs)~\cite{zhao2023survey}, LLM-based generative retrieval has been widely explored for both search~\cite{li2024matching,tay2022transformer,zhuang2022bridging,wang2022neural,zhou2023webultron,bevilacqua2022autoregressive,sun2024learning} and recommendation~\cite{geng2022recommendation,hua2023index,TIGER,LC_Rec,penha2024bridging}. These methods represent items as identifiers and input the user query (for search) or user history (for recommendation) into the LLM to generate the target item.
Identifier designs can be grouped into:
(1) Text-based, using item titles~\cite{dai2023uncovering,liao2024llara} or substrings~\cite{bevilacqua2022autoregressive,li2023multiview};
(2) Non-learnable ID-based, with early methods assigning random IDs~\cite{geng2022recommendation}, and later ones using clustering to encode semantic or collaborative structure~\cite{tay2022transformer,wang2022neural,hua2023index};
(3) Learnable codebook-based, applying techniques like RQ-VAE~\cite{TIGER,LC_Rec} to learn identifiers from semantic or collaborative embeddings.
However, most existing approaches design identifiers tailored to either search or recommendation, focusing solely on semantic or collaborative information. In joint \srcandrec, balancing both is essential for strong performance across tasks.

\begin{figure*}[t]
    \centering
        \includegraphics[width=0.95\textwidth]{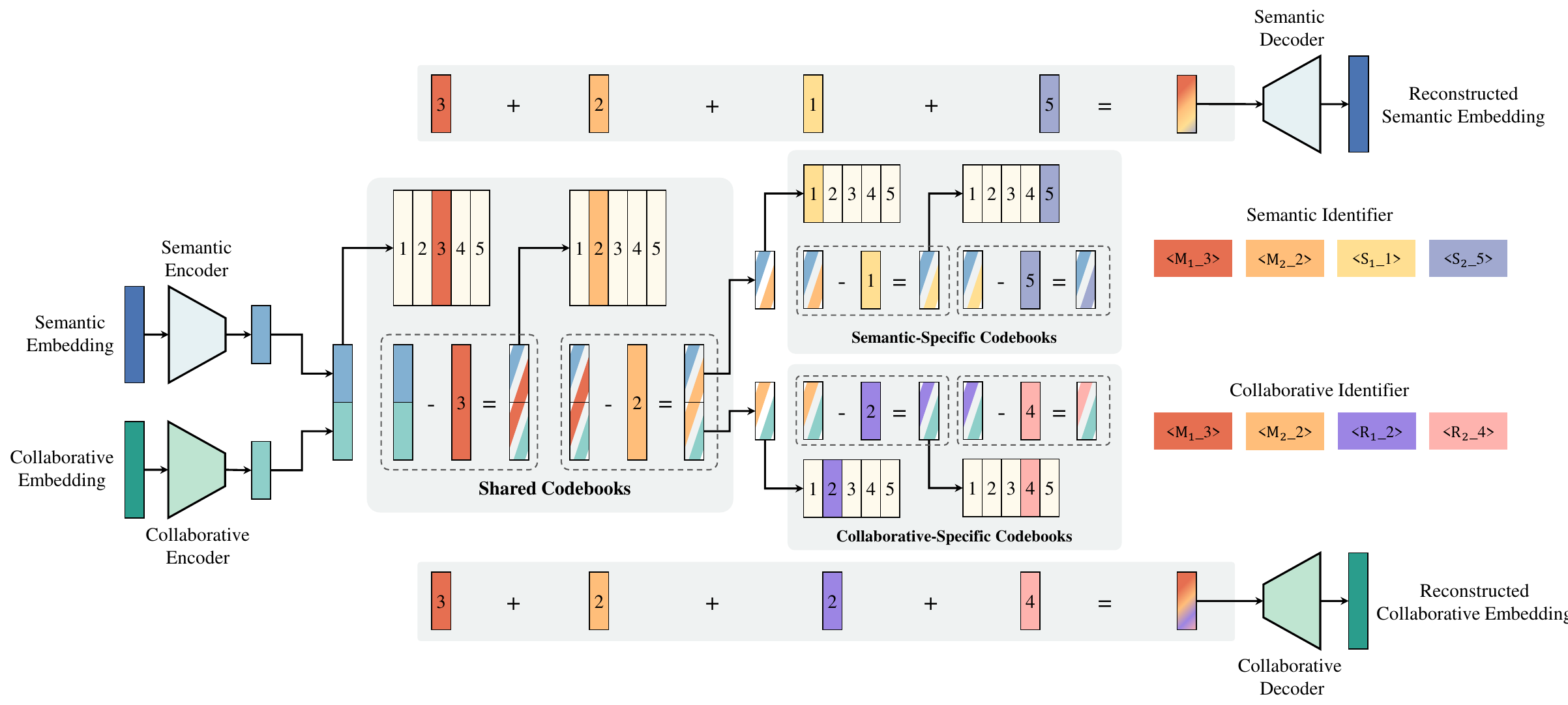}
   \vspace{-10px}
    \caption{
    The joint search and recommendation identifier. We extract the semantic and collaborative embeddings for each item. These two embeddings are first concatenated and passed through the shared codebooks to learn shared codes. Then, the semantic and collaborative embeddings are separately processed through specific codebooks to learn specific codes. Finally, these codes are concatenated to form two identifiers for each item: one for semantics and one for collaboration.
    }
\label{fig:method_identifier}
\vspace{-0.4cm}
\end{figure*}
\section{Our Approach}

This section introduces our proposed method, \ourname.
Section~\ref{sec:formulation} defines the Joint Search and Recommendation task.
Section~\ref{sec:method_identifier} presents the Joint Identifier module, where we design separate semantic and collaborative identifiers to balance the different needs of search and recommendation.
Section~\ref{sec:method_s_r_train} describes task-specific training objectives to help the model capture both types of information.
Finally, Section~\ref{sec:method_train_infer} details the training and inference process of \ourname.

\subsection{Problem Formulation}
\label{sec:formulation}
Let $\mathcal{U},\mathcal{V},\mathcal{Q}$ denote the sets of users, items, and queries, respectively. Each user $u \in \mathcal{U}$ has a chronologically ordered interaction history $S_u = \left[(b_1,x_1),(b_2,x_2),\ldots,(b_N,x_N)\right]$ that includes her historical S\&R behaviors, where $N$ denotes the number of $u$'s historical behaviors. 
$b_i \in \{\text{\recitem}, \text{\srcquery},\text{\srcitem}\}$ represents the type of the $i$-th behavior: \recitem indicates an item clicked by the user after a recommendation, \srcquery represents a query searched by the user, and \srcitem denotes an item clicked by the user after searching a query. $x_i$ denotes the $i$-th behavior:
\begin{equation}
\label{eq:formulation_behavior}
x_i=
\begin{cases}
    v_i, & \text{if}~~b_i=\text{\recitem}~~\text{or}~~b_i=\text{\srcitem}, \\
    q_i, & \text{if}~~b_i=\text{\srcquery},
\end{cases}
\end{equation}
where $v_i \in \mathcal{V}$ denotes the $i$-th interacted item and $q_i \in \mathcal{Q}$ is the $i$-th searched query.
Our goal is to enable the model to understand user interests and predict the next item $v_{N+1}$ for search when $b_{N+1}=\text{\srcitem}$ or recommendation when $b_{N+1}=\text{\recitem}$.

\subsection{Joint Search and Recommendation Identifier}
\label{sec:method_identifier}

This section introduces the design of the joint \srcandrec identifier (Figure~\ref{fig:method_identifier}). We first extract semantic and collaborative embeddings for each item. Using RQ-VAE~\cite{lee2022autoregressive,TIGER,LC_Rec}, we apply both shared and separate codebooks to learn two identifiers per item—one semantic, one collaborative. The identifiers share common parts to capture shared information, while retaining unique parts to reflect task-specific features.

\subsubsection{Embedding Extraction}
For each item $v \in \mathcal{V}$, we can input its textual information, such as the title and description, into a pre-trained retrieval model (e.g., BERT~\cite{devlin-etal-2019-bert}, BGE~\cite{xiao2024c}) to obtain an embedding $\mathbf{v}_s \in \mathbb{R}^{d_s}$ that contains its semantic information. Meanwhile, we can also obtain an embedding $\mathbf{v}_c \in \mathbb{R}^{d_c}$ containing its collaborative information from a pre-trained recommendation model (e.g., SASRec~\cite{SASREC}, BERT4Rec~\cite{BERT4REC}). $d_s$ and $d_c$ represent the dimensions of the semantic and collaborative embeddings, respectively.
We map the semantic and collaborative embeddings to the same-dimensional latent space using two encoders:
\begin{equation}
\label{eq:rq_encoder}
    \mathbf{z}_s = \mathrm{Encoder}_s(\mathbf{v}_s), \quad
    \mathbf{z}_c = \mathrm{Encoder}_c(\mathbf{v}_c), 
\end{equation}
where $\mathrm{z}_s \in \mathbb{R}^d, \mathrm{z}_c \in \mathbb{R}^d$ and $d$ is the dimension of the latent embeddings, $\mathrm{Encoder}_s(\cdot)$ and $\mathrm{Encoder}_c(\cdot)$ are two MLPs (Multilayer~Perceptrons).

\subsubsection{Residual Quantization}
To integrate both semantic and collaborative information, we use $L_m$-level shared codebooks, along with $L_n$-level specific codebooks for semantic and collaborative information, respectively. First, the latent embeddings for semantic and collaborative information, $\mathbf{z}_s$ and $\mathbf{z}_c$, are concatenated to form $\mathbf{r}_0^m=\left[\mathbf{z}_s;\mathbf{z}_c\right] \in \mathbb{R}^{2d}$. This $\mathbf{r}_0^m$ is then passed through the $L_m$-level shared codebooks to obtain the shared codes $I_m$ and the residual embedding $\mathbf{r}_{L_m}^m$.
Then, we extract the semantic part $\mathbf{r}_0^{s}=\mathbf{r}_{L_m}^m\left[1\text{:}d\right] \in \mathbb{R}^{d}$ and the collaborative part $\mathbf{r}_0^{c}=\mathbf{r}_{L_m}^m\left[d\text{:}2d\right] \in \mathbb{R}^{d}$ from $\mathbf{r}_{L_m}^m$, and input them separately into the semantic and collaborative codebooks to learn their specific codes $I_s$ and $I_c$, respectively. 
Finally, the shared and specific codes are concatenated, resulting in two identifiers, $I_{m+s}$ and $I_{m+c}$, for each item.
Next, we will introduce the residual quantization process for both the shared and specific~codebooks.

\vspace{3pt}
\noindent\textbf{$\bullet $ Shared Codebooks.}
We have $L_m$-level shared codebooks. At each level $i\in \{1,2,\ldots,L_m\}$, we have a shared codebook $\mathcal{C}_i^{m} = \{\mathbf{e}_k\}_{k=1}^{K}$, where $K$ is the size of each codebook and $\mathbf{e}_k \in \mathbb{R}^{2d}$ is a learnable code embedding. The residual quantization process for the shared codebooks is as follows:
\begin{equation}
\label{eq:rq_shared}
\begin{aligned}
c_i^m &= \argmin_{k} || \mathbf{r}_{i-1}^m - \mathbf{e}_{k}||^2_2, \quad \mathbf{e}_{k} \in \mathcal{C}_i^m, \\
\mathbf{r}_{i}^m &= \mathbf{r}_{i-1}^m - \mathbf{e}_{c_i^m}, \quad \mathbf{r}_0^m = \left[\mathbf{z}_s;\mathbf{z}_c\right] \in \mathbb{R}^{2d}, \\
\end{aligned}
\end{equation}
where $c_i^m$ is the assigned code from the $i$-th level of the shared codebook. $\mathbf{r}_{i-1}^m$ is the residual from last level.
Through the recursive quantization in Eq.~\eqref{eq:rq_shared}, we can obtain the shared codes $I_m=\left[c_1^m,c_2^m,\ldots,c_{L_m}^m\right]$ and the residual embedding $\mathbf{r}_{L_m}^m$.

\vspace{3pt}
\noindent\textbf{$\bullet $ Specific Codebooks.} We can extract the semantic part $\mathbf{r}_0^{s}=\mathbf{r}_{L_m}^m\left[1\text{:}d\right] \in \mathbb{R}^{d}$ and the collaborative part $\mathbf{r}_0^{c}=\mathbf{r}_{L_m}^m\left[d\text{:}2d\right] \in \mathbb{R}^{d}$ from the residual embedding $\mathbf{r}_{L_m}^m$ outputted by the shared codebooks.
We then pass them separately through the $L_n$-level semantic and collaborative specific codebooks 
$\mathcal{C}_i^{s}$ 
and $\mathcal{C}_i^{c}$, where $i \in \{1,2,\ldots,L_n\}$. 
Please note that, unlike the shared codebook whose code embeddings are $2d$-dimensional, the code embeddings of the specific codebooks are $d$-dimensional.
The residual quantization process for the specific codebooks can be formulated as follows:
\begin{equation}
\label{eq:rq_specific}
\begin{aligned}
&c_i^s = \argmin_{k} || \mathbf{r}_{i-1}^s - \mathbf{e}_{k}||^2_2, \quad \mathbf{e}_{k} \in \mathcal{C}_i^s, \\
&c_i^c = \argmin_{k} || \mathbf{r}_{i-1}^c - \mathbf{e}_{k}||^2_2, \quad \mathbf{e}_{k} \in \mathcal{C}_i^c, \\
&\mathbf{r}_{i}^s = \mathbf{r}_{i-1}^s - \mathbf{e}_{c_i^s},\quad \mathbf{r}_{i}^c = \mathbf{r}_{i-1}^c - \mathbf{e}_{c_i^c},
\end{aligned}
\end{equation}
where $c_i^s$ and $c_i^r$ represent the codes assigned by the $i$-th level semantic-specific and collaborative-specific codebooks, respectively. 
Through the recursive quantization in Eq.~\eqref{eq:rq_specific}, we can obtain the semantic-specific and collaborative-specific codes as follows:
\begin{equation*}
    I_s=\left[c_1^s,c_2^s,\ldots,c_{L_n}^s\right], \quad I_c=\left[c_1^c,c_2^c,\ldots,c_{L_n}^c\right].
\end{equation*}
Finally, by concatenating the shared codes and the specific codes, we can obtain the semantic identifier $I_{m+s}$ and collaborative identifier $I_{m+c}$ for item $v$: 
\begin{equation}
\label{eq:rq_index}
\begin{aligned}
I_{m+s}=\left[c_1^m,c_2^m,\ldots,c_{L_m}^m,c_1^s,c_2^s,\ldots,c_{L_n}^s\right], \\ 
I_{m+c}=\left[c_1^m,c_2^m,\ldots,c_{L_m}^m,c_1^c,c_2^c,\ldots,c_{L_n}^c\right].
\end{aligned}
\end{equation}

\subsubsection{Identifier Training}
After passing through the shared and specific codebooks, we can obtain the semantic and collaborative quantized embeddings as follows:
\begin{equation}
\label{eq:rq_emb}
\hat{\mathbf{z}}_s = \sum_{i=1}^{L_m} \mathbf{e}_{c^m_i}\left[1\text{:}d\right] + \sum_{i=1}^{L_n} \mathbf{e}_{c^s_i}, \quad
\hat{\mathbf{z}}_c = \sum_{i=1}^{L_m} \mathbf{e}_{c^m_i}\left[d\text{:}2d\right] + \sum_{i=1}^{L_n} \mathbf{e}_{c^c_i},
\end{equation}
where $\mathbf{e}_{c^m_i} \in \mathbb{R}^{2d}$ is the code embedding of the shared codebooks, $\mathbf{e}_{c^s_i} \in \mathbb{R}^{d}$ and $\mathbf{e}_{c^c_i} \in \mathbb{R}^{d}$ are the code embeddings of the semantic and collaborative specific codebooks.
The quantized semantic embedding $\hat{\mathbf{z}}_s \in \mathbb{R}^{d}$ and collaborative embedding $\hat{\mathbf{z}}_c \in \mathbb{R}^{d}$ will be used to reconstruct the original semantic and collaborative embeddings, $\mathbf{v}_s$ and $\mathbf{v}_c$:
\begin{equation}
\label{eq:rq_decoder}
    \hat{\mathbf{v}}_{s} = \mathrm{Decoder}_s(\hat{\mathbf{z}}_s), \quad
    \hat{\mathbf{v}}_{c} = \mathrm{Decoder}_c(\hat{\mathbf{z}}_c), 
\end{equation}
where $\mathrm{Decoder}_s(\cdot)$ and $\mathrm{Decoder}_c(\cdot)$ are two MLPs.
We can compute the reconstruction loss used for training the encoder and decoder as follows:
\begin{equation}
\label{eq:rq_recon}
    \mathcal{L}_{\text{Recon}} = ||\mathbf{v}_s - \hat{\mathbf{v}}_{s}||^2_2 + ||\mathbf{v}_c - \hat{\mathbf{v}}_{c}||^2_2. \\
\end{equation}
We can also compute the loss for residual quantization as follows:
\begin{equation}
\label{eq:rq_quan}
\begin{aligned}
&\mathcal{L}_{\text{RQ}}^{m} = \sum_{i=1}^{L_m} ||\mathrm{sg}[\mathbf{r}_{i-1}^m] - \mathbf{e}_{c_i^m}||^2_2 + \alpha||\mathbf{r}_{i-1}^m - \mathrm{sg}[\mathbf{e}_{c_i^m}] ||^2_2,  \\
&\mathcal{L}_{\text{RQ}}^{s} = \sum_{i=1}^{L_n} ||\mathrm{sg}[\mathbf{r}_{i-1}^s] - \mathbf{e}_{c_i^s}||^2_2 + \alpha||\mathbf{r}_{i-1}^s - \mathrm{sg}[\mathbf{e}_{c_i^s}] ||^2_2,  \\
&\mathcal{L}_{\text{RQ}}^{c} = \sum_{i=1}^{L_n} ||\mathrm{sg}[\mathbf{r}_{i-1}^c] - \mathbf{e}_{c_i^c}||^2_2 + \alpha||\mathbf{r}_{i-1}^c - \mathrm{sg}[\mathbf{e}_{c_i^c}] ||^2_2,  \\
&\mathcal{L}_{\text{RQ}}= \mathcal{L}_{\text{RQ}}^{m} + \mathcal{L}_{\text{RQ}}^{s} + \mathcal{L}_{\text{RQ}}^{c}, \\
\end{aligned}
\end{equation}
where $\mathrm{sg}[\cdot]$ denotes the stop-gradient operation and $\alpha$ is a hyper-parameter.
$\mathcal{L}_{\text{RQ}}$ is used to train the code embeddings in both the shared and specific codebooks. Finally, the total loss for training the identifier is as follows:
\begin{equation}
\label{eq:rq_total_loss}
\mathcal{L}_{\text{RQ-VAE}} =\mathcal{L}_{\text{Recon}} + \mathcal{L}_{\text{RQ}}.
\end{equation}

\subsubsection{Behavior-aware Identifier}
\label{sec:behavior_identifier}
After learning the semantic and collaborative identifiers for each item, we can represent each user interaction $(b_i,x_i)$ as shown in Eq.~\eqref{eq:formulation_behavior}. 
To help the model understand different behaviors in the user's interaction history, we prepend a token indicating the behavior type to each interaction's identifier.
For interactions involving items, we prepend the corresponding behavior token to the identifier of each item. For interactions involving queries, we prepend the behavior token to the word sequence of the query.
It can be formulated as follows:
\begin{small}
\begin{equation}
\label{eq:identifier}
\mathrm{ID}(b_i, x_i)=
\begin{cases}
\left[\text{\recitem},c_1^m,c_2^m,\ldots,c_{L_m}^m,c_1^c,c_2^c,\ldots,c_{L_n}^c\right], & \text{if}~~b_i=\text{\recitem}, \\
\left[\text{\srcquery},w_1,w_2,\ldots,w_{|q_i|}\right], & \text{if}~~b_i=\text{\srcquery}, \\
\left[\text{\srcitem},c_1^m,c_2^m,\ldots,c_{L_m}^m,c_1^s,c_2^s,\ldots,c_{L_n}^s\right], & \text{if}~~b_i=\text{\srcitem},
\end{cases}
\end{equation}
\end{small}
where $\left[w_1,w_2,\ldots,w_{|q_i|}\right]$ are the words of query $q_i$. $\mathrm{ID}(\cdot)$ denotes the function for obtaining the identifier of each interaction.

\begin{figure}[t]
    \centering
        \includegraphics[width=0.95\columnwidth]{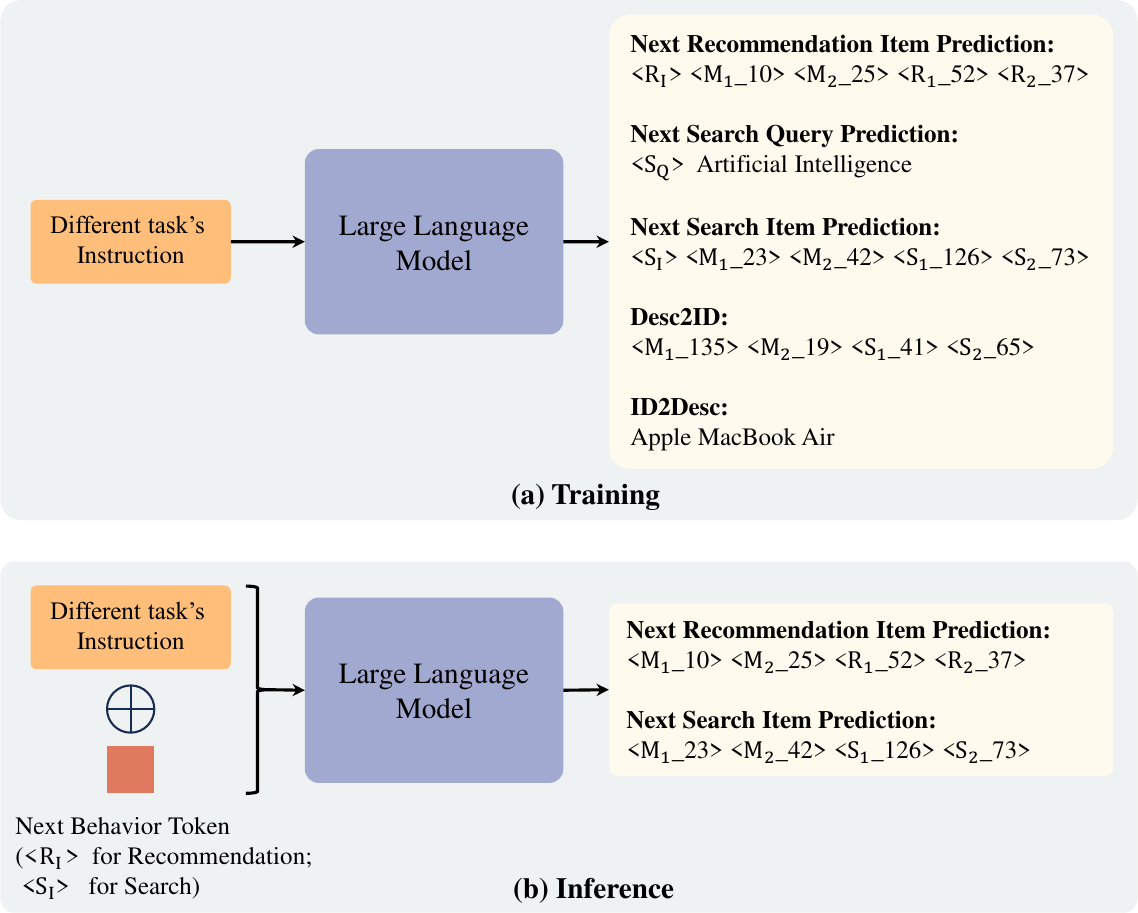}
   \vspace{-8px}
    \caption{
    Training and Inference Process of \ourname. During training, we provide LLM with different instructions to generate corresponding responses. During inference, we append a token at the end of the instruction to indicate the type of behavior to be predicted, enabling the LLM to be applied to either search or recommendation tasks.}
\label{fig:method_model}
\vspace{-0.5cm}
\end{figure}

\subsection{Joint Search and Recommendation Training}
\label{sec:method_s_r_train}

To better adapt the LLM to joint \srcandrec tasks, we design training objectives that help it understand user behaviors and effectively learn both semantic and collaborative identifiers.

\subsubsection{Next Recommendation Item Prediction}
\label{sec:next_rec_item}
To enable the LLM to perform well on the recommendation task, we let it predict the next recommended item. Unlike previous generative recommendation models~\cite{geng2022recommendation,TIGER,LC_Rec} that only use the user's recommendation history, our approach incorporates search history as well. This allows the LLM to better leverage the user's historical information and understand the relationship between \srcandrec behaviors. A sample of the data is shown below:
\begin{tcolorbox}[colframe=black!75!white, title=Next Recommendation Item Prediction]
\textbf{Instruction:} Below is the user's interaction history: \texttt{\srcquery Piano; \srcitem <\sharecodea\_247> <\sharecodeb\_197> <\srccodea\_184> <\srccodeb\_110>; ...; \recitem <\sharecodea\_30> <\sharecodeb\_147> <\reccodea\_247> <\reccodeb\_229>}. 
Please recommend the next item the user is likely to click. \\
\textbf{Response:} \recitem <\sharecodea\_10> <\sharecodeb\_25> <\reccodea\_52> <\reccodeb\_37>
\end{tcolorbox}
\vspace{-0.1cm}
Here, ``<\sharecodea\_10> <\sharecodeb\_25>'' represents the shared semantic and collaborative identifier of the item, ``<\srccodea\_184> <\srccodeb\_110>'' represents the semantic-specific identifier, and ``<\reccodea\_52> <\reccodeb\_37>'' represents the collaborative-specific identifier.

\subsubsection{Next Search Query Prediction}
\label{sec:next_src_query}
Some works focus on query recommendation~\cite{gu2021self,baek2024knowledge,wang2023exploiting}, where they predict the next query a user is likely to search. 
Since our user interaction history also includes search queries, we introduce a task that allows the LLM to predict the user's next intended search query based on their history. This helps the model better understand user search intent and the relationship between \srcandrec behaviors.
A sample of the data for this task is as follows:
\vspace{-0.18cm}
\begin{tcolorbox}[colframe=black!75!white, title=Next Search Query Prediction]
\textbf{Instruction:} Below is the user's interaction history: \texttt{\recitem <\sharecodea\_199> <\sharecodeb\_175> <\reccodea\_1> <\reccodeb\_44>; \recitem <\sharecodea\_209> <\sharecodeb\_235> <\reccodea\_159> <\reccodeb\_80>; ...; \recitem <\sharecodea\_147> <\sharecodeb\_68> <\reccodea\_118> <\reccodeb\_85>}. 
Please predict the next query the user might want to search. \\
\textbf{Response:} \srcquery Artificial Intelligence
\end{tcolorbox}

\subsubsection{Next Search Item Prediction}
\label{sec:next_src_item}
To enable the model to perform well on the search task, we have it predict the next search item. 
Previous generative search models~\cite{tay2022transformer,zhuang2022bridging} only input the user's query into the LLM to predict the target item, which considers only the correlation between the query and the item, without taking the user's preferences into account.
To address this, we include the user's \srcandrec history in the input to reflect their preferences.
A sample of the data for this task is as follows:
\begin{tcolorbox}[colframe=black!75!white, title=Next Search Item Prediction]
\textbf{Instruction:} Below is the user's interaction history: \texttt{\recitem <\sharecodea\_199> <\sharecodeb\_175> <\reccodea\_1> <\reccodeb\_44>; \recitem <\sharecodea\_209> <\sharecodeb\_235> <\reccodea\_159> <\reccodeb\_80>; ...; \recitem <\sharecodea\_147> <\sharecodeb\_68> <\reccodea\_118> <\reccodeb\_85>}.
The user's search query is \texttt{\srcquery Artificial Intelligence}. 
Please predict the next item the user might click. \\
\textbf{Response:} \srcitem <\sharecodea\_23> <\sharecodeb\_42> <\srccodea\_126> <\srccodeb\_73>
\end{tcolorbox}
Here, ``\srcquery Artificial Intelligence'' denotes the query that the user is currently searching for.

\subsubsection{Identifier-Language Alignment} 
\label{sec:id_text_align}
To enhance the LLM's understanding of both the collaborative and semantic identifiers of each item, we designed an identifier-language alignment task. This task enables the LLM to generate a corresponding description based on an item's identifier and, conversely, to generate the appropriate identifier from the item's description.

First, we have the Desc2ID task, which enables the LLM to generate the corresponding item identifier based on its description.
\begin{tcolorbox}[colframe=black!75!white, title=Desc2ID]
\textbf{Instruction:} Using the provided description ``\texttt{Apple MacBook Air}'', predict the corresponding item.  \\
\textbf{Response:} <\sharecodea\_135> <\sharecodeb\_19> <\srccodea\_41> <\srccodeb\_65>
\end{tcolorbox}

Then, we have the ID2Desc task, which enables the LLM to generate the corresponding item description based on its identifier.
\begin{tcolorbox}[colframe=black!75!white, title=ID2Desc]
\textbf{Instruction:} Please provide a description for the item \texttt{<\sharecodea\_135> <\sharecodeb\_19> <\srccodea\_41> <\srccodeb\_65>}. \\
\textbf{Response:} Apple MacBook Air.
\end{tcolorbox}
Please note that for both semantic and collaborative identifiers, we include the Desc2ID and ID2Desc training tasks.
Since the input and output of these two tasks do not involve user history, we do not prepend a token indicating the behavior type to the identifier.

\subsection{Training and Inference}
\label{sec:method_train_infer}
This section introduces how to train the LLM for joint \srcandrec, and how to use the trained LLM during inference to generate the target item for either the search or recommendation task.
The training and inference process of \ourname is shown in Figure~\ref{fig:method_model}.

\subsubsection{Training}
As previously mentioned, each interaction in the user's history is represented as an identifier, allowing us to formulate the task as a sequence-to-sequence problem. We train the model using next token prediction, optimizing the negative log-likelihood of generating the target as follows:
\begin{equation}
\label{eq:loss_ntp}
    \mathcal{L} = -\sum_{t=1}^{T} log~P (y_t|y_{\textless t}, ~ \mathrm{Ins}).
\end{equation}
Here, $y$ represents the behavior-aware identifier of the target to be predicted, as defined in Eq.~\eqref{eq:identifier}. 
$T$ is the length of the identifier of the target item.
$\mathrm{Ins}$ refers to the various instructions described in Section~\ref{sec:method_s_r_train}, which are used as inputs for the LLM.

\subsubsection{Inference}
During training, we train the LLM according to the input-output format described in Section~\ref{sec:method_s_r_train}.
During inference, to apply the LLM to search and recommendation tasks,
we append a behavior token, either ``\srcitem'' for search or ``\recitem'' for recommendation, to the input of the LLM to prompt it to generate the corresponding next item for search or recommendation, respectively. The other tasks mentioned in Section~\ref{sec:method_s_r_train} are used as auxiliary tasks during training to help the model better understand user \srcandrec behaviors.
During generation, to ensure that the items generated by the LLM are within the candidate set, we follow previous works~\cite{hua2023index,LC_Rec} and use constrained beam search.

\subsection{Discussion}
As shown in Table~\ref{tab:comparison}, we compare \ourname with various generative search or recommendation methods in terms of scale (number of parameters), backbone architecture used, and applicable tasks.
\ourname adopts T5-small as its backbone, resulting in a relatively small number of parameters while being capable of serving both \srcandrec tasks. Compared with existing methods, it achieves an optimal balance between efficiency and effectiveness.

In terms of novelty, unlike existing methods that focus solely on either semantic or collaborative information in identifier design, our approach incorporates both the semantic information required for search and the collaborative signals essential for recommendation. This joint consideration helps alleviate the trade-off between \srcandrec.

\begin{table}[t]
    \centering
    \caption{
    Comparison of different generative search or recommendation methods. ``S.'' and  ``R.'' denote search and recommendation respectively.
   }
    \vspace{-5px}
    \label{tab:comparison}
    \resizebox{.98\columnwidth}{!}{
    \begin{tabular}
    {lcc
    cc
    cc}
    \toprule
    \multicolumn{1}{l}{\multirow{2}{*}{Methods}} & 
    \multicolumn{1}{c}{\multirow{2}{*}{Scale}} & 
    \multicolumn{1}{c}{\multirow{2}{*}{Backbone}} & 
    \multicolumn{2}{c}{Task} & 
    \multicolumn{2}{c}{Identifier} \\
    \cmidrule(l){4-5} \cmidrule(l){6-7} 
    \multicolumn{1}{c}{} & \multicolumn{1}{c}{} & \multicolumn{1}{c}{} 
    &S. &R. &Semantic &Collaborative \\
    \midrule
    P5~\cite{geng2022recommendation,hua2023index} &60M/220M &T5-small/T5-base &\textcolor{purple}{\XSolidBrush} & \textcolor{teal}{\CheckmarkBold} &\textcolor{purple}{\XSolidBrush} & \textcolor{teal}{\CheckmarkBold} \\
    TIGER~\cite{TIGER} &60M &T5-small &\textcolor{purple}{\XSolidBrush} & \textcolor{teal}{\CheckmarkBold} &\textcolor{teal}{\CheckmarkBold} &\textcolor{purple}{\XSolidBrush}\\
    LC-Rec~\cite{LC_Rec} &7B &LLaMA &\textcolor{purple}{\XSolidBrush} & \textcolor{teal}{\CheckmarkBold} &\textcolor{teal}{\CheckmarkBold} &\textcolor{purple}{\XSolidBrush}\\
    \hdashline
    DSI-QG\cite{zhuang2022bridging} &220M &T5-base &\textcolor{teal}{\CheckmarkBold} &\textcolor{purple}{\XSolidBrush} &\textcolor{teal}{\CheckmarkBold} &\textcolor{purple}{\XSolidBrush}\\
    WebUltron~\cite{zhou2023webultron} &220M &T5-base &\textcolor{teal}{\CheckmarkBold} &\textcolor{purple}{\XSolidBrush} &\textcolor{teal}{\CheckmarkBold} &\textcolor{purple}{\XSolidBrush}\\
    GenRet~\cite{sun2024learning} &220M &T5-base &\textcolor{teal}{\CheckmarkBold} &\textcolor{purple}{\XSolidBrush} &\textcolor{teal}{\CheckmarkBold} &\textcolor{purple}{\XSolidBrush}\\
    \hdashline
    \ourname (Ours) &60M &T5-small &\textcolor{teal}{\CheckmarkBold} &\textcolor{teal}{\CheckmarkBold} &\textcolor{teal}{\CheckmarkBold} &\textcolor{teal}{\CheckmarkBold}\\
    \bottomrule
    \end{tabular}
    }
    \vspace{-0.3cm}
\end{table}

\section{Experiments}
We conducted experiments to evaluate the performance of \ourname. 

\subsection{Experimental Setup}

\subsubsection{
Dataset
} 
\label{sec:exp:dataset}
We conducted experiments on the following datasets: 
(1)~\textbf{Amazon}\footnote{\url{https://cseweb.ucsd.edu/~jmcauley/datasets/amazon/links.html},~\url{https://github.com/QingyaoAi/Amazon-Product-Search-Datasets}}~\cite{amazon_dataset,amazon_dataset2}: Following previous works~\cite{ai2017learning,ai2019zero,SESRec,UniSAR}, we use the semi-synthetic dataset based on Amazon recommendation data as the public dataset for our experiments.
\footnote{
Please note that 70\% of the items in the ``Kindle Store'' subset used in previous works~\cite{SESRec,UniSAR} lack textual information, so we use the ``Electronics'' subset, where less than 1\% of the items lack text.
}
(2)~\textbf{Commercial}: 
To thoroughly evaluate the effectiveness of \ourname, we collected a dataset from a Chinese commercial app, containing \srcandrec interactions from 10,000 users over two weeks.
For details on data processing and train/validation/test splitting, 
please see the code link.

\subsubsection{Baselines}
In this work, we use the following representative methods as baselines for comparison with \ourname.

First, we compare with the following recommendation models:
(1)~\emph{Sequential Recommendation}:
\textbf{GRU4Rec}~\cite{GRU4REC};
\textbf{SASRec}~\cite{SASREC};
\textbf{FMLP-Rec}~\cite{FMLPREC};
\textbf{LRURec}~\cite{LRURec}.
(2)~\emph{Generative Recommendation}:
\textbf{P5-CID}~\cite{geng2022recommendation,hua2023index};
\textbf{TIGER}~\cite{TIGER};
\textbf{LC-Rec}~\cite{LC_Rec}.
Next, we compare with the following search models:
(1)~\emph{Personalized Search}:
\textbf{QEM}~\cite{ai2019zero};
\textbf{TEM}~\cite{bi2020transformer};
\textbf{CoPPS}~\cite{CoPPS}.
(2)~\emph{Dense Retrieval}:
\textbf{E5}\footnote{\url{https://huggingface.co/intfloat/multilingual-e5-base}}~\cite{wang2024multilingual};
\textbf{BGE}\footnote{\url{https://huggingface.co/BAAI/bge-base-en-v1.5},~\url{https://huggingface.co/BAAI/bge-base-zh-v1.5}}~\cite{xiao2024c}.
(3)~\emph{Generative Retrieval}:
\textbf{DSI-QG}~\cite{zhuang2022bridging};
\textbf{WebUltron}~\cite{zhou2023webultron};
\textbf{GenRet}~\cite{sun2024learning}.
Finally, we compare with the following joint \srcandrec models:
\textbf{JSR}~\cite{JSR};
\textbf{SESRec}~\cite{SESRec};
\textbf{UnifiedSSR}~\cite{xie2024unifiedssr};
\textbf{UniSAR}~\cite{UniSAR}.
For more details on the baselines, 
please see the code link.

\subsubsection{Evaluation Metrics \& 
Implementation Details
}
Following previous works~\cite{FMLPREC,SESRec,UniSAR}, we use ranking metrics including top-$k$ \textit{Hit Ratio} (HR) and top-$k$ \textit{Normalized Discounted Cumulative Gain} (NDCG).
We report the results for $k$ values of \{1, 5, 10\}, and since NDCG@1 is the same as HR@1, we do not report it. 
For more details on the evaluation and model implementation, 
please see the code link.

\begin{table}[t]
    \caption{Statistics of the datasets used in this paper. 
    ``S'' and ``R'' denote search and recommendation, respectively.
    }
    \vspace{-8px}
    \center
     \resizebox{.98\columnwidth}{!}{
        \begin{tabular}{cccccc}
        \toprule
        Dataset & \#Users & \#Items & \#Queries & \#Interaction-R &\#Interaction-S  \\
        \midrule
        Amazon  & 192,403  & 62,883 & 983 & 1,266,903 & 1,081,934 \\
        Commercial  & 10,000 & 782,225 &135,206  &4,286,866  &383,465  \\ 
        \bottomrule
        \end{tabular}}
    \label{tab:dataStatistics}   
   \vspace{-0.3cm}
\end{table}

\begin{table*}[t]
\small
\centering
\caption{
The \emph{recommendation} performance of different methods on the two datasets. 
The best and the second-best methods are highlighted in bold and underlined fonts, respectively.
The improvements over the second-best methods are statistically significant ($t$-test, $p$-value$<0.05$).
Following commonly used settings~\cite{FMLPREC, SESRec, UniSAR}, we pair the ground-truth item with 99 randomly sampled items that the user has not interacted with to form the candidate list.
}
\vspace{-3px}
\label{tab:rec_result}
\resizebox{.98\textwidth}{!}{
\begin{tabular}{
ll
ccccccc
ccccc
}
\toprule
\multicolumn{1}{l}{\multirow{2}{*}{Datasets}} & 
\multicolumn{1}{l}{\multirow{2}{*}{Metrics}} & 
\multicolumn{7}{c}{\textbf{Recommendation}} & 
\multicolumn{5}{c}{\textbf{Joint Search and Recommendation}} \\ 
\cmidrule(l){3-9} \cmidrule(l){10-14}
\multicolumn{1}{c}{}  & \multicolumn{1}{c}{} 
& GRU4Rec & SASRec 
& FMLP-Rec & LRURec & P5-CID & TIGER &LC-Rec &JSR &SESRec &UnifiedSSR & UniSAR & \textbf{\ourname}  \\ 
\midrule
\multirow{5} * {Amazon}
&HR@1 &0.0440 &0.0544 &0.0534 &0.0544 &0.0881 &\underline{0.1073} &0.1063 &0.0657 &0.0627 &0.0477 &0.0680 &\textbf{0.1261} \\
&HR@5 &0.1716 &0.1887 &0.1898 &0.1890 &0.1874 &0.2046 &0.1973 &0.2075 &0.2083 &0.1667 &\underline{0.2171} &\textbf{0.2228} \\
&HR@10 &0.2884 &0.2992 &0.3041 &0.3001 &0.2790 &0.2852 &0.2760 &0.3188 &\underline{0.3209} &0.2707 &\textbf{0.3319} &0.3063 \\
&NDCG@5 &0.1074 &0.1216 &0.1217 &0.1218 &0.1380 &\underline{0.1565} &0.1522 &0.1371 &0.1359 &0.1071 &0.1432 &\textbf{0.1748} \\
&NDCG@10 &0.1449 &0.1571 &0.1584 &0.1575 &0.1674 &\underline{0.1824} &0.1774 &0.1729 &0.1721 &0.1405 &0.1802 &\textbf{0.2015} \\
\hline
\multirow{5} * {Commercial}
&HR@1 &0.1022 &0.1519 &0.1442 &0.1363 &\underline{0.2843} &0.2630 &0.2703 &0.1576 &0.1890 &0.1515 &0.2214 &\textbf{0.2997} \\
&HR@5 &0.2526 &0.2812 &0.2711 &0.2637 &\underline{0.3305} &0.3013 &0.3001 &0.2685 &0.2845 &0.2844 &0.3228 &\textbf{0.3496} \\
&HR@10 &0.3527 &0.3716 &0.3584 &0.3525 &0.3830 &0.3448 &0.3333 &0.3529 &0.3690 &0.3870 &\textbf{0.4056} &\underline{0.4031} \\
&NDCG@5 &0.1787 &0.2179 &0.2093 &0.2021 &\underline{0.3072} &0.2819 &0.2849 &0.2142 &0.2370 &0.2195 &0.2727 &\textbf{0.3241}\\
&NDCG@10 &0.2110 &0.2470 &0.2373 &0.2306 &\underline{0.3240} &0.2958 &0.2955 &0.2413 &0.2641 &0.2524 &0.2993 &\textbf{0.3411}\\
\bottomrule
\end{tabular} 
}
\vspace{-0.3cm}
\end{table*}

\subsection{Overall Performance}
Table~\ref{tab:rec_result} and Table~\ref{tab:src_result} show the \srcandrec results on two datasets, respectively. From the results, we can observe that:

\noindent\textbf{$\bullet $}~Firstly, it can be seen that compared to existing search or recommendation models, \ourname achieves state-of-the-art results. This validates the effectiveness of \ourname in alleviating the trade-off between \srcandrec through generative retrieval, by designing joint identifiers and training tasks for both tasks.

\noindent\textbf{$\bullet $}~~Secondly, we can observe that most joint \srcandrec methods (e.g., JSR, UniSAR, \ourname) outperform traditional methods that using only item IDs, such as sequential recommendation (e.g., SASRec, FMLP-Rec) and personalized search methods (e.g., QEM, TEM, CoPPS). This demonstrates the advantages of jointly modeling of \srcandrec, as it enhances the performance of both tasks.

\noindent\textbf{$\bullet $}~Thirdly, it can be observed that for search, dense retrieval (e.g., E5, BGE) and generative retrieval (e.g., GenRet, \ourname) methods that rely on semantic information outperform personalized search models (e.g., QEM, TEM, CoPPS) that rely solely on ID information. This also confirms that for search, semantic information is more important than collaborative information.

\begin{table*}[h!]
\small
\centering
\caption{
The \emph{search} performance of different methods on the two datasets. 
Since search relies on semantic relevance, previous works~\cite{xie2024unifiedssr,UniSAR} that randomly sample negatives often produce overly easy examples, leading to inflated performance and poor model differentiation. To address this, we follow prior personalized search methods~\cite{ahmad2018multi,deng2022improving} and use BM25~\cite{robertson2009probabilistic} to retrieve 99 harder negatives, forming a candidate list with the positive sample for more accurate evaluation.
}
\vspace{-3px}
\label{tab:src_result}
\resizebox{.98\textwidth}{!}{
\begin{tabular}{
ll
cccccccc
cccc
}
\toprule
\multicolumn{1}{l}{\multirow{2}{*}{Datasets}} & 
\multicolumn{1}{l}{\multirow{2}{*}{Metrics}} & 
\multicolumn{8}{c}{\textbf{Search}} & 
\multicolumn{4}{c}{\textbf{Joint Search and Recommendation}} \\ 
\cmidrule(l){3-10} \cmidrule(l){11-14}
\multicolumn{1}{c}{}  & \multicolumn{1}{c}{} 
& QEM & TEM & CoPPS 
& E5
& BGE & DSI-QG & WebUltron &GenRet &JSR &UnifiedSSR & UniSAR & \textbf{\ourname}  \\ 
\midrule
\multirow{5} * {Amazon}
&HR@1 &0.1512 &0.0839 &0.0943 &0.3289 &0.4030 &0.3558 &0.3432 &\underline{0.4173} &0.0835 &0.0799 &0.1122 &\textbf{0.5262} \\
&HR@5 &0.3101 &0.3471 &0.3380 &0.5945 &0.6264 &0.5848 &0.5464 &\underline{0.6513} &0.2407 &0.2476 &0.3129 &\textbf{0.7529} \\
&HR@10 &0.4657 &0.5181 &0.4909 &0.7203 &\underline{0.7475} &0.6897 &0.6216 &0.7339 &0.3463 &0.3614 &0.4333 &\textbf{0.8217} \\
&NDCG@5 &0.2311 &0.2173 &0.2154 &0.4662 &0.5219 &0.4764 &0.4507 &\underline{0.5399} &0.1623 &0.1662 &0.2143 &\textbf{0.6485} \\
&NDCG@10 &0.2809 &0.2722 &0.2647 &0.5069 &0.5613 &0.5103 &0.4748 &\underline{0.5667} &0.1962 &0.2028 &0.2533 &\textbf{0.6710} \\
\hline
\multirow{5} * {Commercial}
&HR@1 &0.0311 &0.0328 &0.0265 &\textbf{0.1277} &0.1267 &0.1016 &0.0804 &0.1171 &0.0273 &0.0119 &0.0511 &\underline{0.1249} \\
&HR@5 &0.0870 &0.1106 &0.0998 &0.3108 &0.3184 &0.2831 &0.2619 &\underline{0.3320} &0.1202 &0.0470 &0.1810 &\textbf{0.3655} \\
&HR@10 &0.1539 &0.1925 &0.1792 &0.4044 &0.4194 &0.4132 &0.3992 &\underline{0.4666} &0.2137 &0.0873 &0.3231 &\textbf{0.5250} \\
&NDCG@5 &0.0586 &0.0715 &0.0626 &0.2230 &0.2258 &0.1940 &0.1721 &\underline{0.2273} &0.0728 &0.0292 &0.1144 &\textbf{0.2472} \\
&NDCG@10 &0.0799 &0.0977 &0.0880 &0.2533 &0.2584 &0.2359 &0.2164 &\underline{0.2708} &0.1026 &0.0420 &0.1597 &\textbf{0.2987} \\
\bottomrule
\end{tabular} 
}
\vspace{-0.3cm}
\end{table*}

\subsection{Ablation Study}
We conducted ablation study on the Commercial dataset to validate the effectiveness of the various training tasks in \ourname, as shown in Table~\ref{tab:ablation_result}.

\textbf{Impact of Behavior Token.}
As shown in Section~\ref{sec:behavior_identifier}, we prepended a token indicating the type of behavior to the identifier of each user interaction, enabling the LLM to recognize different behavior types. To evaluate its impact, we removed this behavior token, as shown in Table~\ref{tab:ablation_result} (``w/o Behavior Token''). The results indicate that removing the behavior token degrades performance, validating that adding this token helps the LLM better understand the relationship between user \srcandrec behaviors.

\textbf{Next Recommendation Item Prediction (NRIP).}
As shown in Section~\ref{sec:next_rec_item}, we incorporated the training task ``Next Recommendation Item Prediction'' (NRIP), which enables the LLM to predict the next item to recommend based on user history. To evaluate its impact, we removed this task, as shown in Table~\ref{tab:ablation_result} (``w/o NRIP''). The results demonstrate that removing this task significantly degrades recommendation performance and slightly reduces search performance, highlighting the importance of NRIP. 
Additionally, this demonstrates that recommendation training tasks can enhance search performance, verifying that recommendation can benefit~search.

\textbf{Next Search Query Prediction (NSQP).}
We included the training task ``Next Search Query Prediction'' (NSQP) to enable the LLM to better understand user intent by predicting the next query a user might want to search, as described in Section~\ref{sec:next_src_query}. To evaluate its impact, we observed the results after removing this task, as shown in Table~\ref{tab:ablation_result} (``w/o NSQP''). The results indicate that removing this task significantly degrades search performance and also affects recommendation performance, demonstrating that NSQP helps the model better understand user search intent.

\textbf{Next Search Item Prediction (NSIP).}
In Section~\ref{sec:next_src_item}, we introduced the training task ``Next Search Item Prediction'' (NSIP), which allows the LLM to predict the next item a user might click based on their history and input query. We analyzed the impact of this task, as shown in Table~\ref{tab:ablation_result} (``w/o NSIP''). The results indicate that removing this task significantly degrades search performance, while also slightly affecting recommendation performance. This demonstrates the importance of NSIP for search and further highlights that search training tasks can enhance recommendation performance, validating that search can assist recommendation.

\textbf{Identifier-Language Alignment.}
In Section~\ref{sec:id_text_align}, we introduced two tasks, Desc2ID and ID2Desc, for identifier-language alignment, which help the LLM better understand the semantic and collaborative identifiers of each item. We observed the impact of removing these two tasks, as shown in Table~\ref{tab:ablation_result} (w/o ``Desc2ID'' and w/o ``ID2Desc''). It can be seen that removing these tasks leads to a decrease in both \srcandrec performance, indicating the effectiveness of these tasks in helping the LLM better understand item identifiers.

\begin{table}[t!]
    \small
    \caption{
    Ablation study on the Commercial dataset, where ``w/o'' denotes the removal of the corresponding module in~\ourname.}
    \vspace{-8px}
    \label{tab:ablation_result}
    \renewcommand{\arraystretch}{1.2}
 \resizebox{.95\columnwidth}{!}{
 \begin{tabular}
 {lcccc}
    \toprule
    \multicolumn{1}{l }{\multirow{2}{*}{Model}} & 
 \multicolumn{2}{c}{Recommendation} & 
 \multicolumn{2}{c}{Search}  	\\ 
 \cmidrule(l){2-3} \cmidrule(l){4-5}
    \multicolumn{1}{c}{}   
&HR@5 &NDCG@5
&HR@5 &NDCG@5  \\ 
 \midrule
\textbf{\ourname} &\textbf{0.3496} &\textbf{0.3241} &\textbf{0.3655} &\textbf{0.2472} \\
\midrule
w/o Behavior Token &0.3430 &0.3193 &0.3298 &0.2224 \\ 
\hdashline
w/o NRIP &0.0665 &0.0392 &0.3456 &0.2342 \\
w/o NSQP &0.3401 &0.3163 &0.3089 &0.2053 \\
w/o NSIP &0.3390 &0.3152 &0.1668 &0.1113 \\
\hdashline
w/o Desc2ID &0.3416 &0.3188 &0.3355 &0.2278 \\
w/o ID2Desc &0.3458 &0.3220 &0.3398 &0.2308 \\
\bottomrule
\end{tabular}
 } 
 \vspace{-0.5cm}
\end{table}

\subsection{Experimental Analysis}
We conducted further experiments on the Commercial dataset to analyze the effectiveness of different modules in \ourname.

\begin{figure}[t]
     \centering
     \subfigure[Recommendation Performance]{
        \label{fig:rec_id_result}
        \includegraphics[width=0.475\columnwidth]{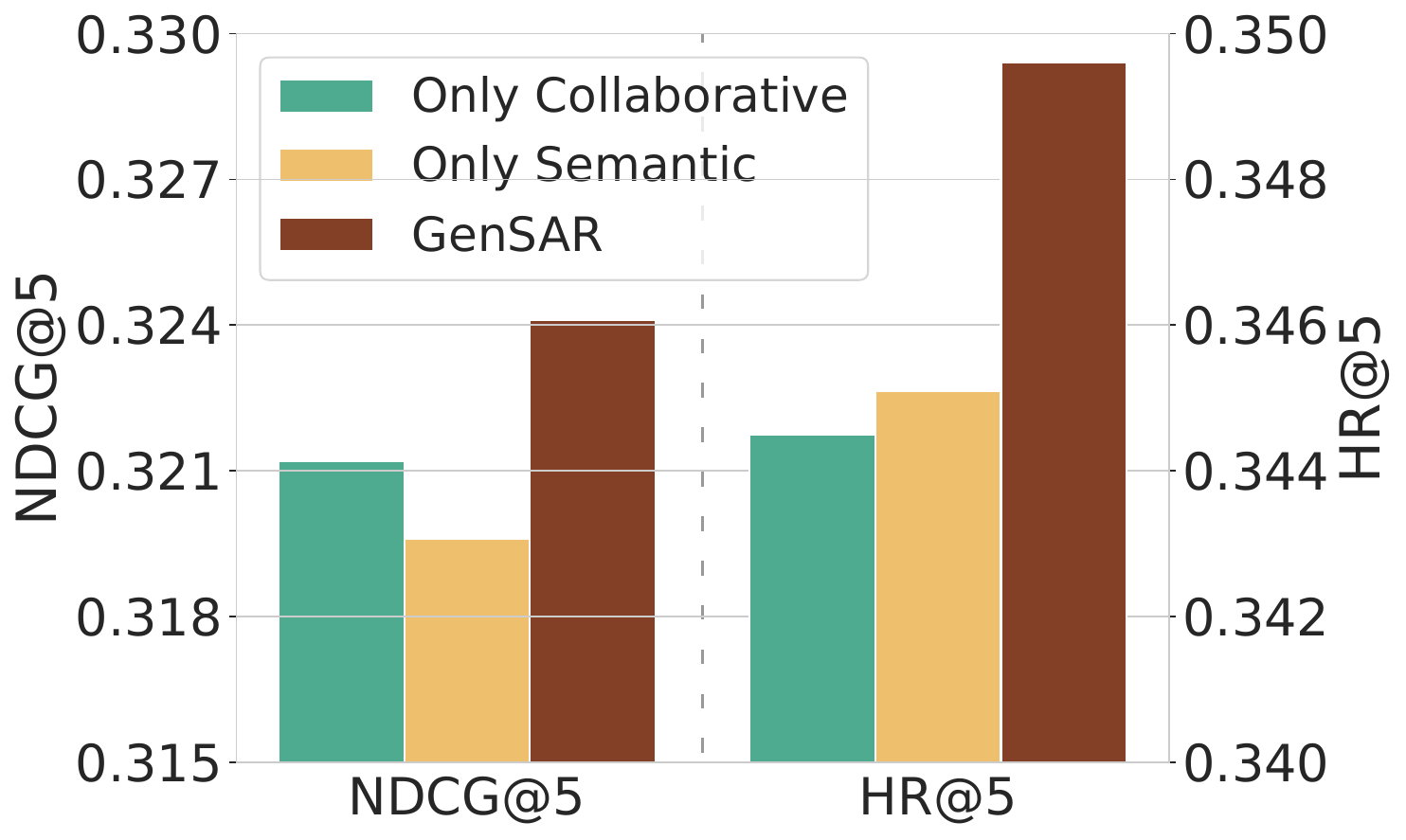}
     }
    \subfigure[Search Performance]{
        \label{fig:src_id_result}
        \includegraphics[width=0.475\columnwidth]{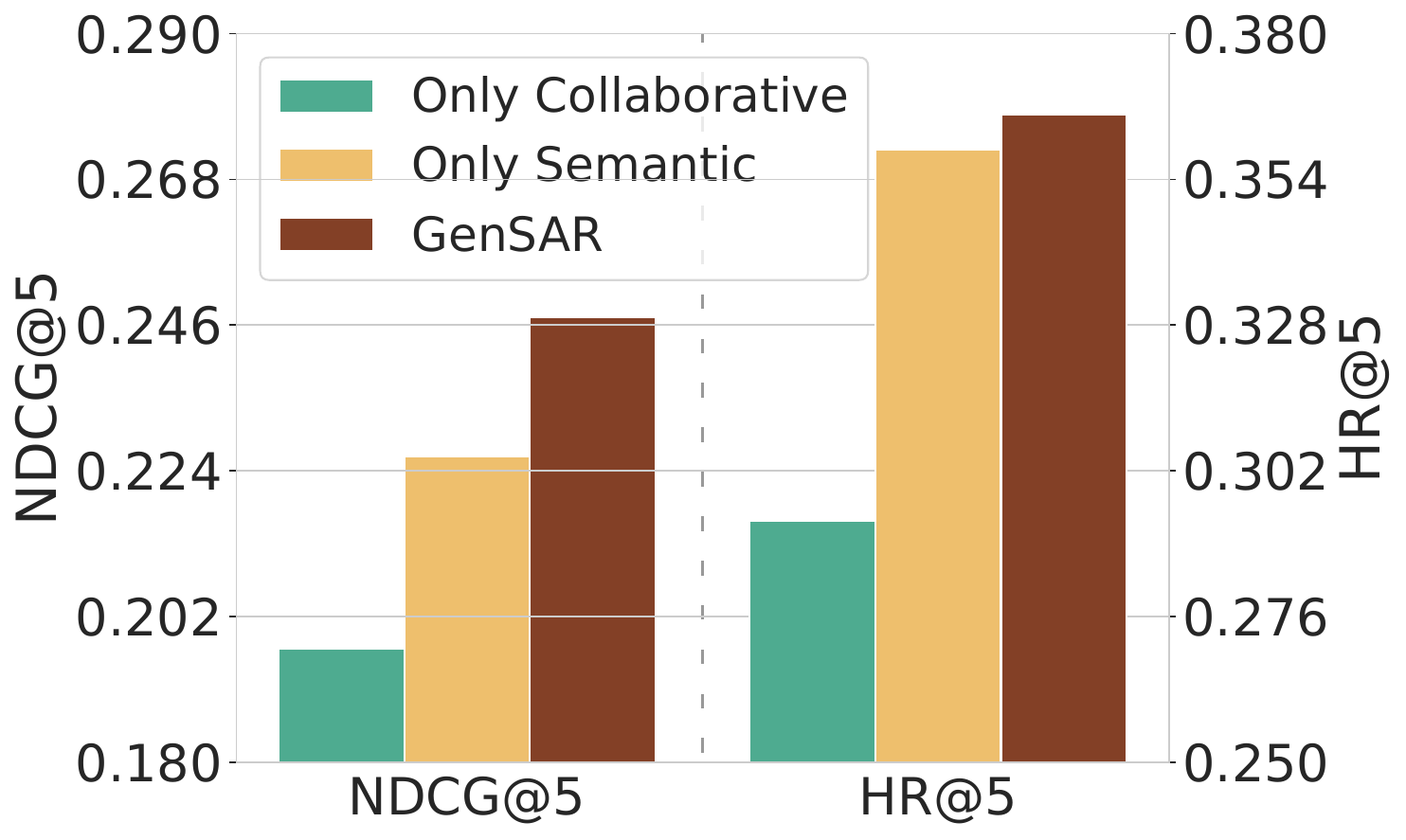}
     }
     \vspace{-8px}
     \caption{
     Performance of \ourname using different identifiers.
     }
     \label{fig:id_result}
     \vspace{-0.3cm}
\end{figure}

\begin{figure}
    \centering
    \includegraphics[width=0.95\columnwidth]{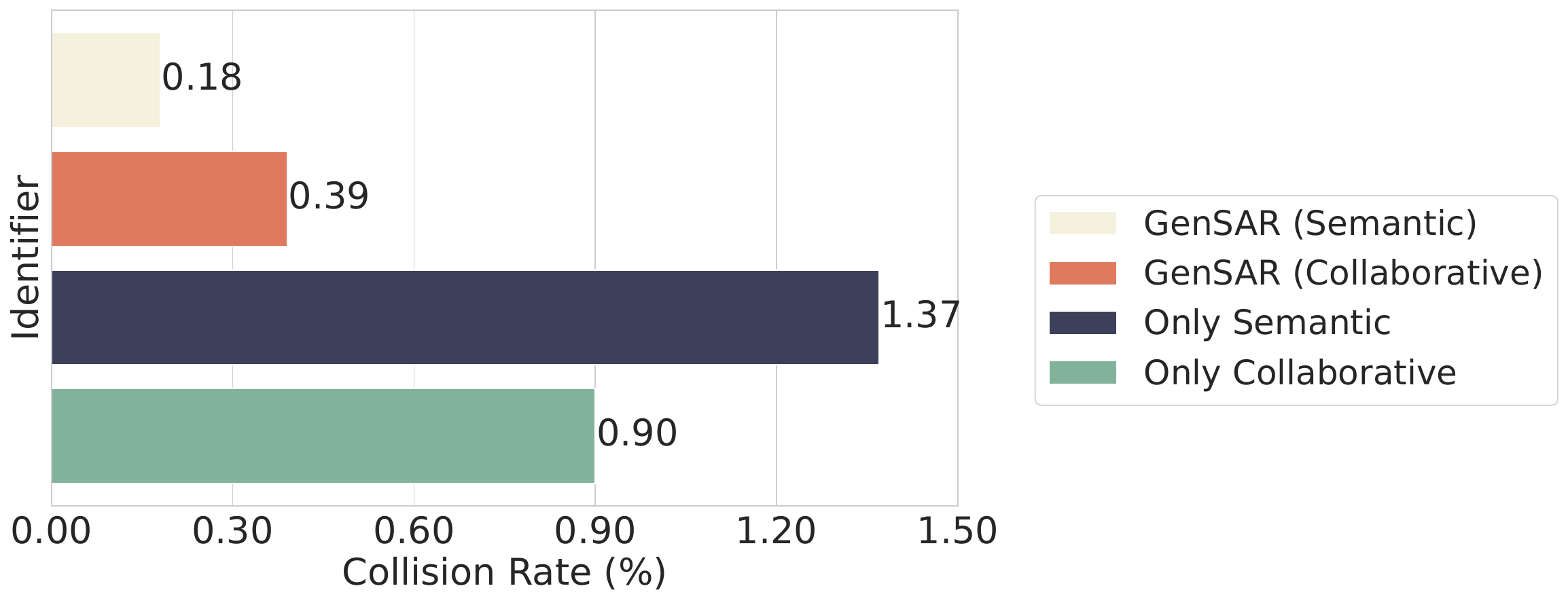}
   \vspace{-8px}
    \caption{
    Collision rate of different identifiers.
    }
    \label{fig:collision_rate}
    \vspace{-0.3cm}
\end{figure}

\subsubsection{Impact of Different Identifier}
To balance the semantic information needed for search and the collaborative information needed for recommendation, we designed the joint \srcandrec identifier in Section~\ref{sec:method_identifier}. To validate its effectiveness, we compared it with identifiers learned directly from semantic embeddings or collaborative embeddings using RQ-VAE~\cite{TIGER,LC_Rec}, as shown in Figure~\ref{fig:id_result}. 
``Only Collaborative'' represents using only collaborative embeddings, while ``Only Semantic'' represents using only semantic embeddings. The results show that identifiers derived solely from semantic or collaborative information lead to degraded performance. Furthermore, using only collaborative information results in worse search performance, which aligns with the fact that search relies more on semantic information.

\subsubsection{Collision Rate of Different Identifier}
Additionally, we analyzed the advantages of different identifiers from the perspective of collision rate. 
The formula for calculating the collision rate is as follows:
\begin{small}
\begin{equation*}
    \mathrm{Collision~Rate} = 1- \frac{\#~\mathrm{Unique~Identifier}}{\#~\mathrm{Unique~Item}},
\end{equation*}
\end{small}
where $\#~\mathrm{Unique~Identifier}$ represents the number of unique identifiers, and $\#~\mathrm{Unique~Item}$ represents the number of unique items.
Since RQ-VAE does not guarantee a unique identifier for each item during the learning process, collisions may occur where different items share the same identifier~\cite{TIGER,LC_Rec}. A higher collision rate can negatively impact the model's performance.
From Figure~\ref{fig:collision_rate}, it can be observed that the two identifiers assigned to each item in \ourname, incorporating both semantic and collaborative information, have a lower collision rate of 0.18\% and 0.39\%, respectively. In contrast, identifiers derived solely from semantic embeddings or collaborative embeddings exhibit higher collision rates of 1.37\% and 0.90\%, respectively. This further validates the advantage of the identifiers in \ourname, as their lower collision rate enables the model to achieve better~performance.

\subsubsection{Impact of Hyper-parameters}
\begin{figure}[t]
     \centering
     \subfigure[Recommendation Performance]{
        \label{fig:rec_shared_code}
        \includegraphics[width=0.475\columnwidth]{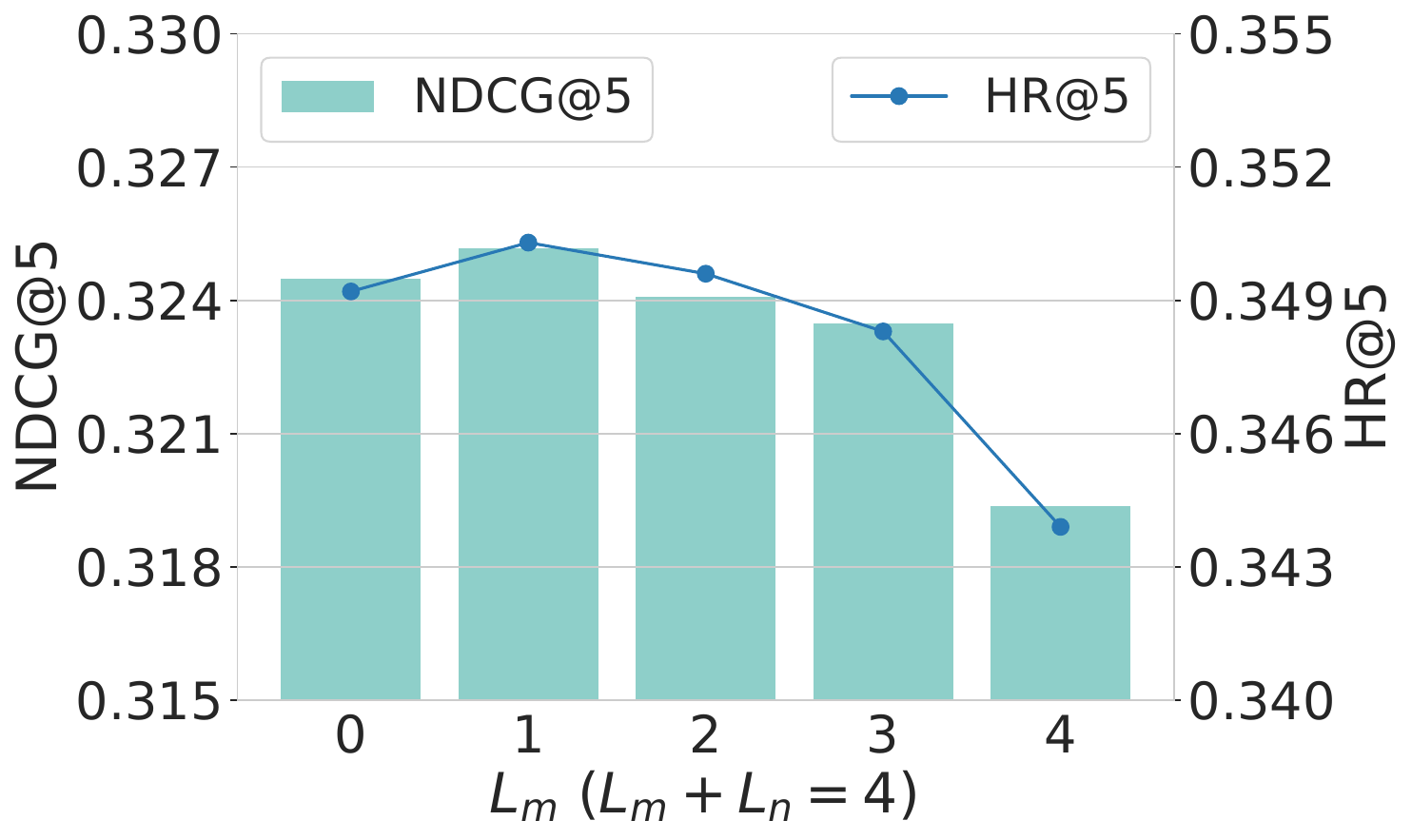}
     }
    \subfigure[Search Performance]{
        \label{fig:src_shared_code}
        \includegraphics[width=0.475\columnwidth]{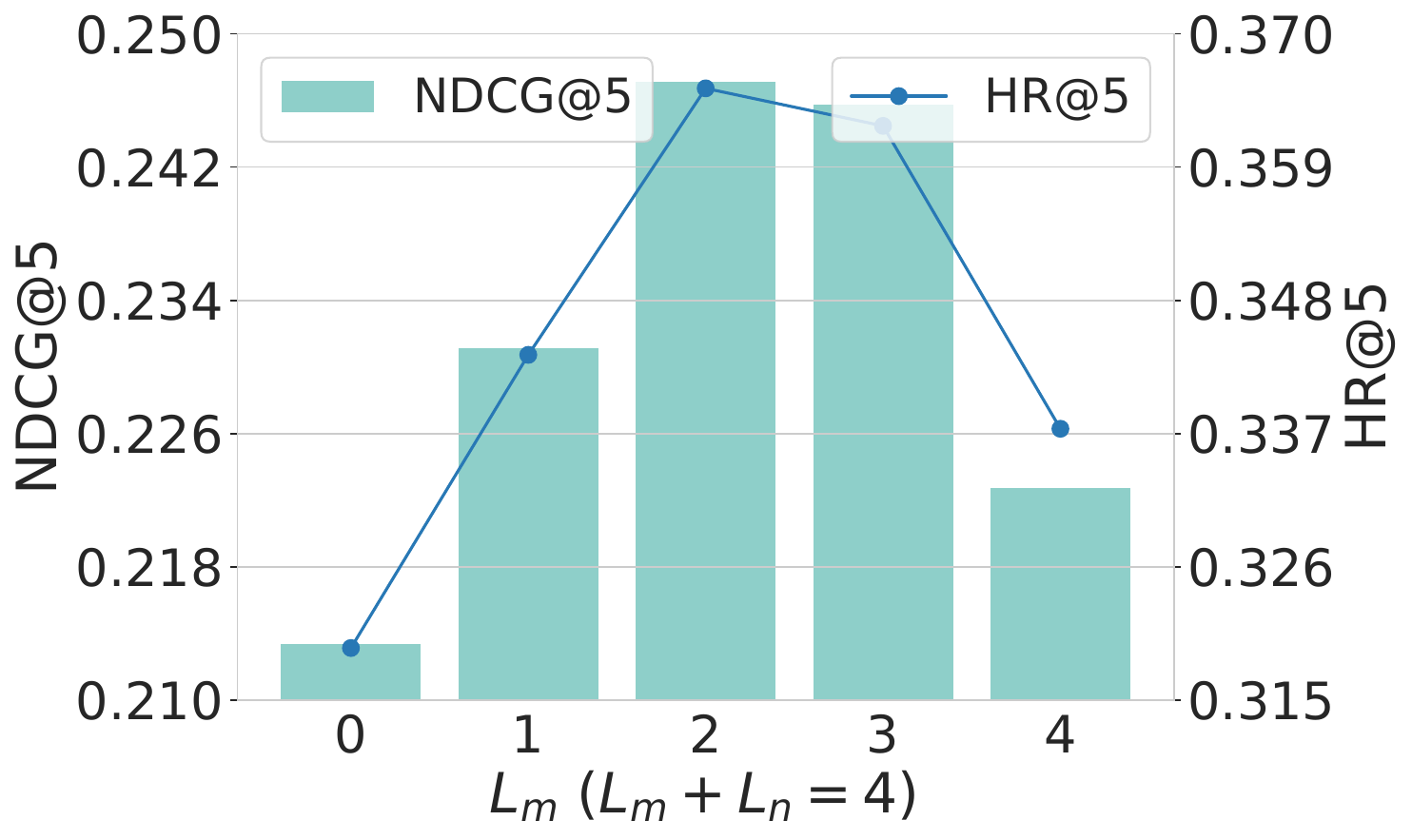}
     }
     \vspace{-8px}
     \caption{
     Performance under different numbers of shared codebooks $L_m$. We fix $L_m+L_n=4$ and vary $L_m$ to observe the~results.
     }
     \label{fig:shared_code}
     \vspace{-0.3cm}
\end{figure}

As described in Section~\ref{sec:method_identifier}, we have $L_m$-level shared codebooks and $L_n$-level specific codebooks. Here, we analyze the impact of the number of shared and specific codebooks ($L_m$ and $L_n$) on the results, as shown in Figure~\ref{fig:shared_code}. We fix $L_m + L_n=4$ and observe the results. It can be seen that having too few ($L_m=1$) or too many ($L_m=3$) shared codebooks fails to achieve strong performance in both \srcandrec. This indicates that $L_m$ needs to be properly set so that the identifier can capture both the shared information between semantics and collaboration as well as their specific characteristics. Only in this way can we achieve better performance in both \srcandrec.

\begin{figure}[t]
     \centering
     \subfigure[Recommendation Performance]{
        \label{fig:rec_id_len}
        \includegraphics[width=0.475\columnwidth]{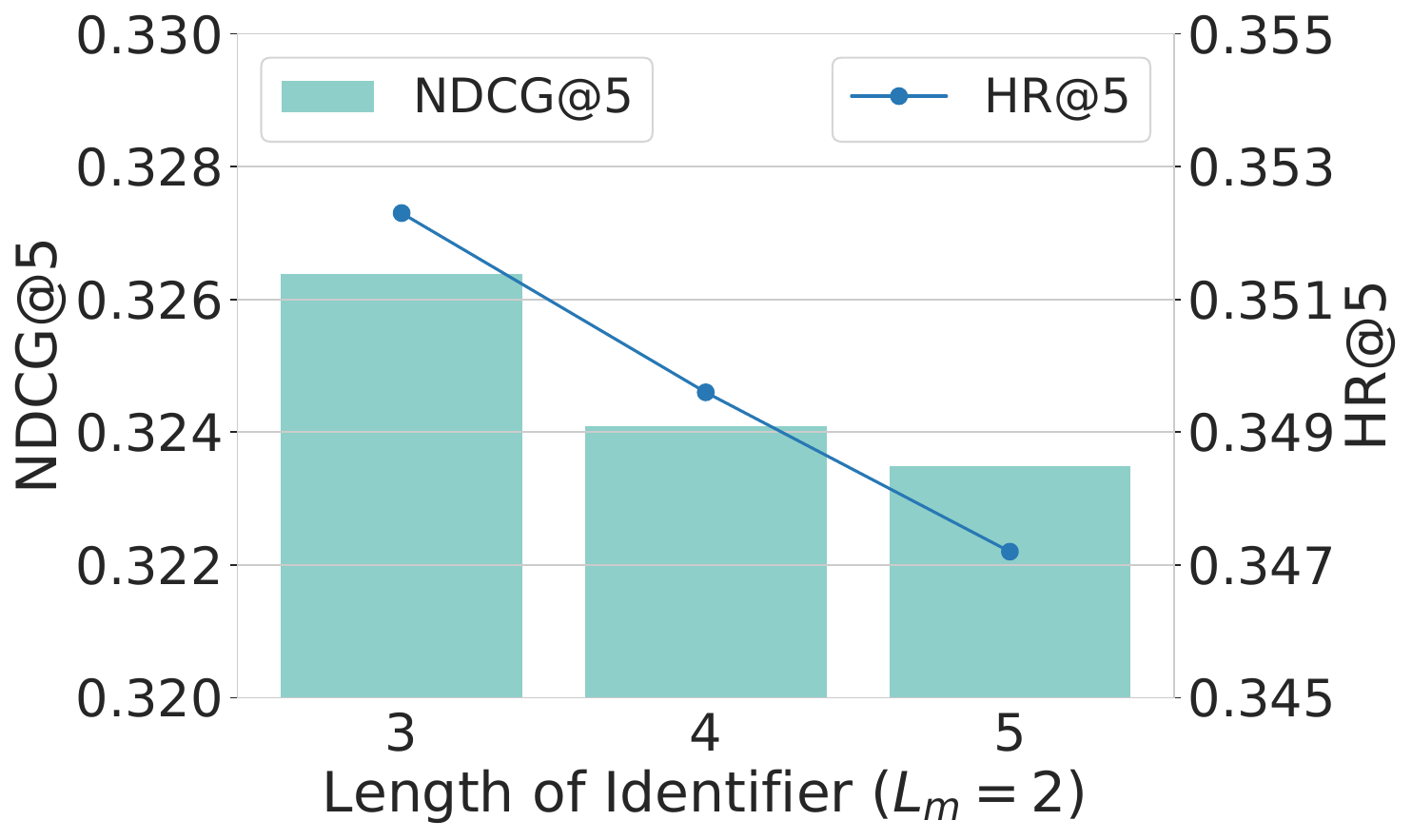}
     }
    \subfigure[Search Performance]{
        \label{fig:src_id_len}
        \includegraphics[width=0.475\columnwidth]{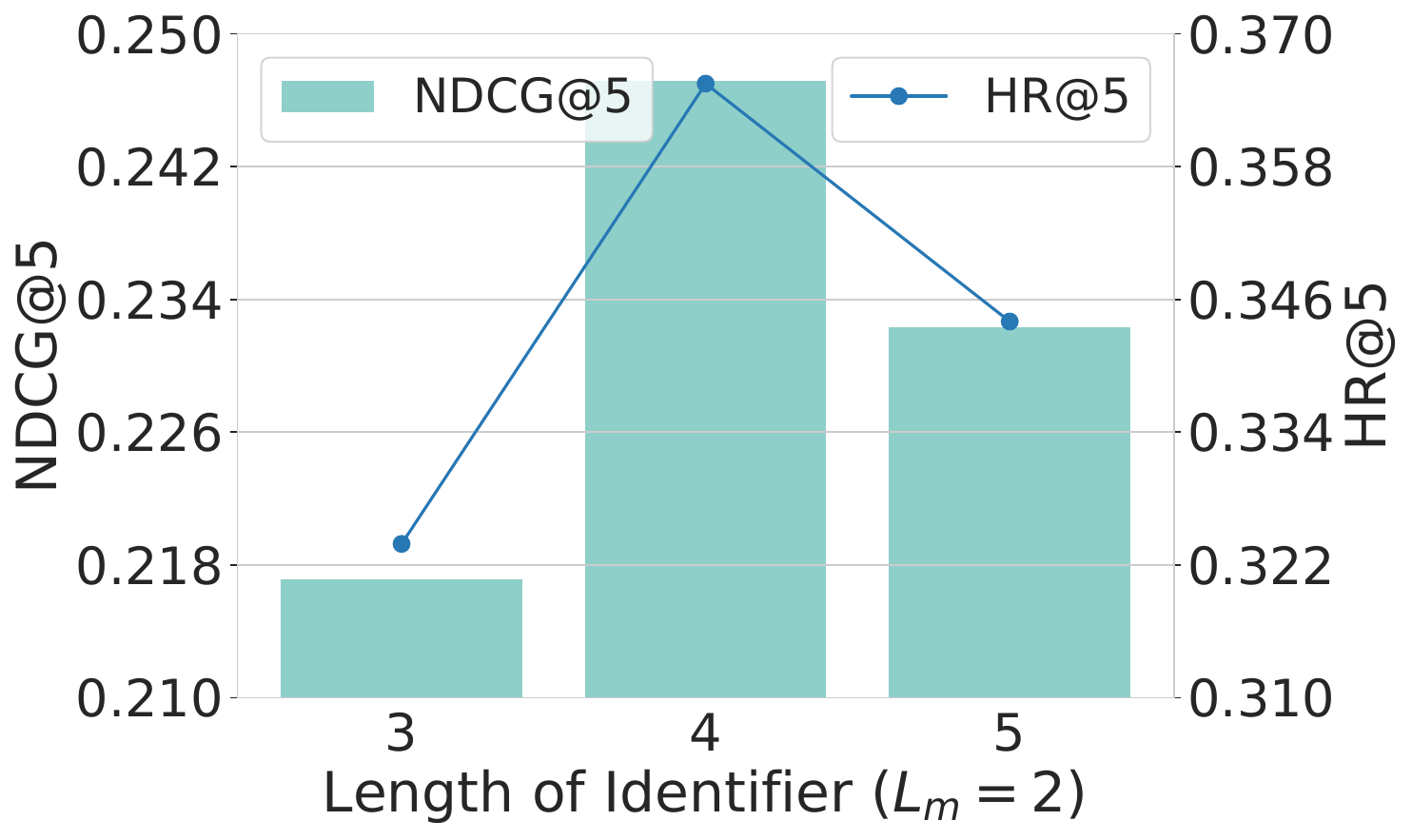}
     }
     \vspace{-8px}
     \caption{
     Performance under different length of the identifier. We fix $L_m=2$ and vary $L_n$ to adjust the identifier length.
     }
     \label{fig:id_len}
     \vspace{-0.3cm}
\end{figure}

Additionally, we analyzed the impact of identifier length on performance, as shown in Figure~\ref{fig:id_len}. We fix $L_m=2$ and vary $L_n$ to adjust the identifier length and observe the results. It can be seen that both shorter ($L_m+L_n=3$) and longer ($L_m+L_n=5$) identifiers lead to performance degradation. This is because, when the identifier is too short, the identifiers learned through RQ-VAE are more prone to collisions, resulting in a higher collision rate and making it difficult for the model to distinguish between different items. On the other hand, when the identifier is too long, the model requires more decoding steps during item generation, leading to accumulated errors and ultimately deteriorating performance. 
Therefore, it is essential to properly set the identifier length to achieve better performance.

\section{Conclusion}
In this paper, we propose \ourname, which unifies balanced search and recommendation through generative retrieval to alleviate the trade-off between the two tasks and improve their performance. 
To balance the semantic information required for search and the collaborative information needed for recommendation, we design the joint \srcandrec identifier and different training tasks.
First, we learn two identifiers for each item to represent semantic and collaborative information, respectively. These identifiers share a common part to capture the information shared between semantics and collaboration while retaining distinct parts to preserve specific information. Second, we design different training tasks to help the model better understand the requirements of \srcandrec tasks. 
We also validate the effectiveness of \ourname through extensive experiments.

\bibliographystyle{ACM-Reference-Format}
\bibliography{ref}


\begin{thebibliography}{59}


\ifx \showCODEN    \undefined \def \showCODEN     #1{\unskip}     \fi
\ifx \showDOI      \undefined \def \showDOI       #1{#1}\fi
\ifx \showISBNx    \undefined \def \showISBNx     #1{\unskip}     \fi
\ifx \showISBNxiii \undefined \def \showISBNxiii  #1{\unskip}     \fi
\ifx \showISSN     \undefined \def \showISSN      #1{\unskip}     \fi
\ifx \showLCCN     \undefined \def \showLCCN      #1{\unskip}     \fi
\ifx \shownote     \undefined \def \shownote      #1{#1}          \fi
\ifx \showarticletitle \undefined \def \showarticletitle #1{#1}   \fi
\ifx \showURL      \undefined \def \showURL       {\relax}        \fi
\providecommand\bibfield[2]{#2}
\providecommand\bibinfo[2]{#2}
\providecommand\natexlab[1]{#1}
\providecommand\showeprint[2][]{arXiv:#2}

\bibitem[Ahmad et~al\mbox{.}(2018)]%
        {ahmad2018multi}
\bibfield{author}{\bibinfo{person}{Wasi~Uddin Ahmad}, \bibinfo{person}{Kai-Wei Chang}, {and} \bibinfo{person}{Hongning Wang}.} \bibinfo{year}{2018}\natexlab{}.
\newblock \showarticletitle{Multi-task learning for document ranking and query suggestion}. In \bibinfo{booktitle}{\emph{International conference on learning representations}}.
\newblock


\bibitem[Ai et~al\mbox{.}(2019)]%
        {ai2019zero}
\bibfield{author}{\bibinfo{person}{Qingyao Ai}, \bibinfo{person}{Daniel~N Hill}, \bibinfo{person}{SVN Vishwanathan}, {and} \bibinfo{person}{W~Bruce Croft}.} \bibinfo{year}{2019}\natexlab{}.
\newblock \showarticletitle{A zero attention model for personalized product search}. In \bibinfo{booktitle}{\emph{Proceedings of the 28th ACM International Conference on Information and Knowledge Management}}. \bibinfo{pages}{379--388}.
\newblock


\bibitem[Ai et~al\mbox{.}(2017)]%
        {ai2017learning}
\bibfield{author}{\bibinfo{person}{Qingyao Ai}, \bibinfo{person}{Yongfeng Zhang}, \bibinfo{person}{Keping Bi}, \bibinfo{person}{Xu Chen}, {and} \bibinfo{person}{W~Bruce Croft}.} \bibinfo{year}{2017}\natexlab{}.
\newblock \showarticletitle{Learning a hierarchical embedding model for personalized product search}. In \bibinfo{booktitle}{\emph{Proceedings of the 40th International ACM SIGIR Conference on Research and Development in Information Retrieval}}. \bibinfo{pages}{645--654}.
\newblock


\bibitem[Baek et~al\mbox{.}(2024)]%
        {baek2024knowledge}
\bibfield{author}{\bibinfo{person}{Jinheon Baek}, \bibinfo{person}{Nirupama Chandrasekaran}, \bibinfo{person}{Silviu Cucerzan}, \bibinfo{person}{Allen Herring}, {and} \bibinfo{person}{Sujay~Kumar Jauhar}.} \bibinfo{year}{2024}\natexlab{}.
\newblock \showarticletitle{Knowledge-augmented large language models for personalized contextual query suggestion}. In \bibinfo{booktitle}{\emph{Proceedings of the ACM on Web Conference 2024}}. \bibinfo{pages}{3355--3366}.
\newblock


\bibitem[Bevilacqua et~al\mbox{.}(2022)]%
        {bevilacqua2022autoregressive}
\bibfield{author}{\bibinfo{person}{Michele Bevilacqua}, \bibinfo{person}{Giuseppe Ottaviano}, \bibinfo{person}{Patrick Lewis}, \bibinfo{person}{Scott Yih}, \bibinfo{person}{Sebastian Riedel}, {and} \bibinfo{person}{Fabio Petroni}.} \bibinfo{year}{2022}\natexlab{}.
\newblock \showarticletitle{Autoregressive search engines: Generating substrings as document identifiers}.
\newblock \bibinfo{journal}{\emph{Advances in Neural Information Processing Systems}}  \bibinfo{volume}{35} (\bibinfo{year}{2022}), \bibinfo{pages}{31668--31683}.
\newblock


\bibitem[Bi et~al\mbox{.}(2020)]%
        {bi2020transformer}
\bibfield{author}{\bibinfo{person}{Keping Bi}, \bibinfo{person}{Qingyao Ai}, {and} \bibinfo{person}{W~Bruce Croft}.} \bibinfo{year}{2020}\natexlab{}.
\newblock \showarticletitle{A transformer-based embedding model for personalized product search}. In \bibinfo{booktitle}{\emph{Proceedings of the 43rd International ACM SIGIR Conference on Research and Development in Information Retrieval}}. \bibinfo{pages}{1521--1524}.
\newblock


\bibitem[Dai et~al\mbox{.}(2023a)]%
        {CoPPS}
\bibfield{author}{\bibinfo{person}{Shitong Dai}, \bibinfo{person}{Jiongnan Liu}, \bibinfo{person}{Zhicheng Dou}, \bibinfo{person}{Haonan Wang}, \bibinfo{person}{Lin Liu}, \bibinfo{person}{Bo Long}, {and} \bibinfo{person}{Ji{-}Rong Wen}.} \bibinfo{year}{2023}\natexlab{a}.
\newblock \showarticletitle{Contrastive Learning for User Sequence Representation in Personalized Product Search}. In \bibinfo{booktitle}{\emph{Proceedings of the 29th {ACM} {SIGKDD} Conference on Knowledge Discovery and Data Mining, {KDD} 2023, Long Beach, CA, USA, August 6-10, 2023}}. \bibinfo{publisher}{{ACM}}, \bibinfo{pages}{380--389}.
\newblock


\bibitem[Dai et~al\mbox{.}(2023b)]%
        {dai2023uncovering}
\bibfield{author}{\bibinfo{person}{Sunhao Dai}, \bibinfo{person}{Ninglu Shao}, \bibinfo{person}{Haiyuan Zhao}, \bibinfo{person}{Weijie Yu}, \bibinfo{person}{Zihua Si}, \bibinfo{person}{Chen Xu}, \bibinfo{person}{Zhongxiang Sun}, \bibinfo{person}{Xiao Zhang}, {and} \bibinfo{person}{Jun Xu}.} \bibinfo{year}{2023}\natexlab{b}.
\newblock \showarticletitle{Uncovering chatgpt’s capabilities in recommender systems}. In \bibinfo{booktitle}{\emph{Proceedings of the 17th ACM Conference on Recommender Systems}}. \bibinfo{pages}{1126--1132}.
\newblock


\bibitem[Deng et~al\mbox{.}(2022)]%
        {deng2022improving}
\bibfield{author}{\bibinfo{person}{Chenlong Deng}, \bibinfo{person}{Yujia Zhou}, {and} \bibinfo{person}{Zhicheng Dou}.} \bibinfo{year}{2022}\natexlab{}.
\newblock \showarticletitle{Improving personalized search with dual-feedback network}. In \bibinfo{booktitle}{\emph{Proceedings of the fifteenth ACM international conference on web search and data mining}}. \bibinfo{pages}{210--218}.
\newblock


\bibitem[Devlin et~al\mbox{.}(2019)]%
        {devlin-etal-2019-bert}
\bibfield{author}{\bibinfo{person}{Jacob Devlin}, \bibinfo{person}{Ming-Wei Chang}, \bibinfo{person}{Kenton Lee}, {and} \bibinfo{person}{Kristina Toutanova}.} \bibinfo{year}{2019}\natexlab{}.
\newblock \showarticletitle{{BERT}: Pre-training of Deep Bidirectional Transformers for Language Understanding}. In \bibinfo{booktitle}{\emph{Proceedings of the 2019 Conference of the North {A}merican Chapter of the Association for Computational Linguistics: Human Language Technologies, Volume 1 (Long and Short Papers)}}.
\newblock


\bibitem[Geng et~al\mbox{.}(2022)]%
        {geng2022recommendation}
\bibfield{author}{\bibinfo{person}{Shijie Geng}, \bibinfo{person}{Shuchang Liu}, \bibinfo{person}{Zuohui Fu}, \bibinfo{person}{Yingqiang Ge}, {and} \bibinfo{person}{Yongfeng Zhang}.} \bibinfo{year}{2022}\natexlab{}.
\newblock \showarticletitle{Recommendation as language processing (rlp): A unified pretrain, personalized prompt \& predict paradigm (p5)}. In \bibinfo{booktitle}{\emph{Proceedings of the 16th ACM Conference on Recommender Systems}}. \bibinfo{pages}{299--315}.
\newblock


\bibitem[Gu et~al\mbox{.}(2021)]%
        {gu2021self}
\bibfield{author}{\bibinfo{person}{Yulong Gu}, \bibinfo{person}{Wentian Bao}, \bibinfo{person}{Dan Ou}, \bibinfo{person}{Xiang Li}, \bibinfo{person}{Baoliang Cui}, \bibinfo{person}{Biyu Ma}, \bibinfo{person}{Haikuan Huang}, \bibinfo{person}{Qingwen Liu}, {and} \bibinfo{person}{Xiaoyi Zeng}.} \bibinfo{year}{2021}\natexlab{}.
\newblock \showarticletitle{Self-supervised learning on users' spontaneous behaviors for multi-scenario ranking in e-commerce}. In \bibinfo{booktitle}{\emph{Proceedings of the 30th ACM International Conference on Information \& Knowledge Management}}. \bibinfo{pages}{3828--3837}.
\newblock


\bibitem[He and McAuley(2016)]%
        {amazon_dataset}
\bibfield{author}{\bibinfo{person}{Ruining He} {and} \bibinfo{person}{Julian McAuley}.} \bibinfo{year}{2016}\natexlab{}.
\newblock \showarticletitle{Ups and downs: Modeling the visual evolution of fashion trends with one-class collaborative filtering}. In \bibinfo{booktitle}{\emph{proceedings of the 25th international conference on world wide web}}. \bibinfo{pages}{507--517}.
\newblock


\bibitem[He et~al\mbox{.}(2017)]%
        {he2017neural}
\bibfield{author}{\bibinfo{person}{Xiangnan He}, \bibinfo{person}{Lizi Liao}, \bibinfo{person}{Hanwang Zhang}, \bibinfo{person}{Liqiang Nie}, \bibinfo{person}{Xia Hu}, {and} \bibinfo{person}{Tat-Seng Chua}.} \bibinfo{year}{2017}\natexlab{}.
\newblock \showarticletitle{Neural collaborative filtering}. In \bibinfo{booktitle}{\emph{Proceedings of the 26th international conference on world wide web}}. \bibinfo{pages}{173--182}.
\newblock


\bibitem[He et~al\mbox{.}(2022)]%
        {Query_SeqRec}
\bibfield{author}{\bibinfo{person}{Zhankui He}, \bibinfo{person}{Handong Zhao}, \bibinfo{person}{Zhaowen Wang}, \bibinfo{person}{Zhe Lin}, \bibinfo{person}{Ajinkya Kale}, {and} \bibinfo{person}{Julian Mcauley}.} \bibinfo{year}{2022}\natexlab{}.
\newblock \showarticletitle{Query-Aware Sequential Recommendation}. In \bibinfo{booktitle}{\emph{Proceedings of the 31st ACM International Conference on Information \&amp; Knowledge Management}} (Atlanta, GA, USA) \emph{(\bibinfo{series}{CIKM '22})}. \bibinfo{publisher}{Association for Computing Machinery}, \bibinfo{address}{New York, NY, USA}, \bibinfo{pages}{4019–4023}.
\newblock
\showISBNx{9781450392365}


\bibitem[Hidasi et~al\mbox{.}(2016)]%
        {GRU4REC}
\bibfield{author}{\bibinfo{person}{Bal{\'{a}}zs Hidasi}, \bibinfo{person}{Alexandros Karatzoglou}, \bibinfo{person}{Linas Baltrunas}, {and} \bibinfo{person}{Domonkos Tikk}.} \bibinfo{year}{2016}\natexlab{}.
\newblock \showarticletitle{Session-based Recommendations with Recurrent Neural Networks}. In \bibinfo{booktitle}{\emph{4th International Conference on Learning Representations, {ICLR} 2016, San Juan, Puerto Rico, May 2-4, 2016, Conference Track Proceedings}}, \bibfield{editor}{\bibinfo{person}{Yoshua Bengio} {and} \bibinfo{person}{Yann LeCun}} (Eds.).
\newblock


\bibitem[Hua et~al\mbox{.}(2023)]%
        {hua2023index}
\bibfield{author}{\bibinfo{person}{Wenyue Hua}, \bibinfo{person}{Shuyuan Xu}, \bibinfo{person}{Yingqiang Ge}, {and} \bibinfo{person}{Yongfeng Zhang}.} \bibinfo{year}{2023}\natexlab{}.
\newblock \showarticletitle{How to index item ids for recommendation foundation models}. In \bibinfo{booktitle}{\emph{Proceedings of the Annual International ACM SIGIR Conference on Research and Development in Information Retrieval in the Asia Pacific Region}}. \bibinfo{pages}{195--204}.
\newblock


\bibitem[Izacard et~al\mbox{.}(2021)]%
        {izacard2021unsupervised}
\bibfield{author}{\bibinfo{person}{Gautier Izacard}, \bibinfo{person}{Mathilde Caron}, \bibinfo{person}{Lucas Hosseini}, \bibinfo{person}{Sebastian Riedel}, \bibinfo{person}{Piotr Bojanowski}, \bibinfo{person}{Armand Joulin}, {and} \bibinfo{person}{Edouard Grave}.} \bibinfo{year}{2021}\natexlab{}.
\newblock \showarticletitle{Unsupervised dense information retrieval with contrastive learning}.
\newblock \bibinfo{journal}{\emph{arXiv preprint arXiv:2112.09118}} (\bibinfo{year}{2021}).
\newblock


\bibitem[Kang and McAuley(2018)]%
        {SASREC}
\bibfield{author}{\bibinfo{person}{Wang-Cheng Kang} {and} \bibinfo{person}{Julian McAuley}.} \bibinfo{year}{2018}\natexlab{}.
\newblock \showarticletitle{Self-attentive sequential recommendation}. In \bibinfo{booktitle}{\emph{2018 IEEE International Conference on Data Mining (ICDM)}}. IEEE, \bibinfo{pages}{197--206}.
\newblock


\bibitem[Lee et~al\mbox{.}(2022)]%
        {lee2022autoregressive}
\bibfield{author}{\bibinfo{person}{Doyup Lee}, \bibinfo{person}{Chiheon Kim}, \bibinfo{person}{Saehoon Kim}, \bibinfo{person}{Minsu Cho}, {and} \bibinfo{person}{Wook-Shin Han}.} \bibinfo{year}{2022}\natexlab{}.
\newblock \showarticletitle{Autoregressive image generation using residual quantization}. In \bibinfo{booktitle}{\emph{Proceedings of the IEEE/CVF Conference on Computer Vision and Pattern Recognition}}. \bibinfo{pages}{11523--11532}.
\newblock


\bibitem[Li et~al\mbox{.}(2024)]%
        {li2024matching}
\bibfield{author}{\bibinfo{person}{Xiaoxi Li}, \bibinfo{person}{Jiajie Jin}, \bibinfo{person}{Yujia Zhou}, \bibinfo{person}{Yuyao Zhang}, \bibinfo{person}{Peitian Zhang}, \bibinfo{person}{Yutao Zhu}, {and} \bibinfo{person}{Zhicheng Dou}.} \bibinfo{year}{2024}\natexlab{}.
\newblock \showarticletitle{From matching to generation: A survey on generative information retrieval}.
\newblock \bibinfo{journal}{\emph{arXiv preprint arXiv:2404.14851}} (\bibinfo{year}{2024}).
\newblock


\bibitem[Li et~al\mbox{.}(2023)]%
        {li2023multiview}
\bibfield{author}{\bibinfo{person}{Yongqi Li}, \bibinfo{person}{Nan Yang}, \bibinfo{person}{Liang Wang}, \bibinfo{person}{Furu Wei}, {and} \bibinfo{person}{Wenjie Li}.} \bibinfo{year}{2023}\natexlab{}.
\newblock \showarticletitle{Multiview Identifiers Enhanced Generative Retrieval}. In \bibinfo{booktitle}{\emph{Proceedings of the 61st Annual Meeting of the Association for Computational Linguistics (Volume 1: Long Papers)}}. \bibinfo{pages}{6636--6648}.
\newblock


\bibitem[Liao et~al\mbox{.}(2024)]%
        {liao2024llara}
\bibfield{author}{\bibinfo{person}{Jiayi Liao}, \bibinfo{person}{Sihang Li}, \bibinfo{person}{Zhengyi Yang}, \bibinfo{person}{Jiancan Wu}, \bibinfo{person}{Yancheng Yuan}, \bibinfo{person}{Xiang Wang}, {and} \bibinfo{person}{Xiangnan He}.} \bibinfo{year}{2024}\natexlab{}.
\newblock \showarticletitle{Llara: Large language-recommendation assistant}. In \bibinfo{booktitle}{\emph{Proceedings of the 47th International ACM SIGIR Conference on Research and Development in Information Retrieval}}. \bibinfo{pages}{1785--1795}.
\newblock


\bibitem[McAuley et~al\mbox{.}(2015)]%
        {amazon_dataset2}
\bibfield{author}{\bibinfo{person}{Julian McAuley}, \bibinfo{person}{Christopher Targett}, \bibinfo{person}{Qinfeng Shi}, {and} \bibinfo{person}{Anton Van Den~Hengel}.} \bibinfo{year}{2015}\natexlab{}.
\newblock \showarticletitle{Image-based recommendations on styles and substitutes}. In \bibinfo{booktitle}{\emph{Proceedings of the 38th international ACM SIGIR conference on research and development in information retrieval}}. \bibinfo{pages}{43--52}.
\newblock


\bibitem[Penha et~al\mbox{.}(2024)]%
        {penha2024bridging}
\bibfield{author}{\bibinfo{person}{Gustavo Penha}, \bibinfo{person}{Ali Vardasbi}, \bibinfo{person}{Enrico Palumbo}, \bibinfo{person}{Marco De~Nadai}, {and} \bibinfo{person}{Hugues Bouchard}.} \bibinfo{year}{2024}\natexlab{}.
\newblock \showarticletitle{Bridging Search and Recommendation in Generative Retrieval: Does One Task Help the Other?}. In \bibinfo{booktitle}{\emph{Proceedings of the 18th ACM Conference on Recommender Systems}}. \bibinfo{pages}{340--349}.
\newblock


\bibitem[Rajput et~al\mbox{.}(2023)]%
        {TIGER}
\bibfield{author}{\bibinfo{person}{Shashank Rajput}, \bibinfo{person}{Nikhil Mehta}, \bibinfo{person}{Anima Singh}, \bibinfo{person}{Raghunandan Hulikal~Keshavan}, \bibinfo{person}{Trung Vu}, \bibinfo{person}{Lukasz Heldt}, \bibinfo{person}{Lichan Hong}, \bibinfo{person}{Yi Tay}, \bibinfo{person}{Vinh Tran}, \bibinfo{person}{Jonah Samost}, {et~al\mbox{.}}} \bibinfo{year}{2023}\natexlab{}.
\newblock \showarticletitle{Recommender systems with generative retrieval}.
\newblock \bibinfo{journal}{\emph{Advances in Neural Information Processing Systems}}  \bibinfo{volume}{36} (\bibinfo{year}{2023}), \bibinfo{pages}{10299--10315}.
\newblock


\bibitem[Robertson et~al\mbox{.}(2009)]%
        {robertson2009probabilistic}
\bibfield{author}{\bibinfo{person}{Stephen Robertson}, \bibinfo{person}{Hugo Zaragoza}, {et~al\mbox{.}}} \bibinfo{year}{2009}\natexlab{}.
\newblock \showarticletitle{The probabilistic relevance framework: BM25 and beyond}.
\newblock \bibinfo{journal}{\emph{Foundations and Trends{\textregistered} in Information Retrieval}} \bibinfo{volume}{3}, \bibinfo{number}{4} (\bibinfo{year}{2009}), \bibinfo{pages}{333--389}.
\newblock


\bibitem[Shen et~al\mbox{.}(2024)]%
        {shen2024survey}
\bibfield{author}{\bibinfo{person}{Chenglei Shen}, \bibinfo{person}{Xiao Zhang}, \bibinfo{person}{Teng Shi}, \bibinfo{person}{Changshuo Zhang}, \bibinfo{person}{Guofu Xie}, {and} \bibinfo{person}{Jun Xu}.} \bibinfo{year}{2024}\natexlab{}.
\newblock \showarticletitle{A survey of controllable learning: Methods and applications in information retrieval}.
\newblock \bibinfo{journal}{\emph{arXiv preprint arXiv:2407.06083}} (\bibinfo{year}{2024}).
\newblock


\bibitem[Shi et~al\mbox{.}(2024)]%
        {UniSAR}
\bibfield{author}{\bibinfo{person}{Teng Shi}, \bibinfo{person}{Zihua Si}, \bibinfo{person}{Jun Xu}, \bibinfo{person}{Xiao Zhang}, \bibinfo{person}{Xiaoxue Zang}, \bibinfo{person}{Kai Zheng}, \bibinfo{person}{Dewei Leng}, \bibinfo{person}{Yanan Niu}, {and} \bibinfo{person}{Yang Song}.} \bibinfo{year}{2024}\natexlab{}.
\newblock \showarticletitle{UniSAR: Modeling User Transition Behaviors between Search and Recommendation}. In \bibinfo{booktitle}{\emph{Proceedings of the 47th International ACM SIGIR Conference on Research and Development in Information Retrieval}}. \bibinfo{pages}{1029--1039}.
\newblock


\bibitem[Si et~al\mbox{.}(2022)]%
        {IV4REC}
\bibfield{author}{\bibinfo{person}{Zihua Si}, \bibinfo{person}{Xueran Han}, \bibinfo{person}{Xiao Zhang}, \bibinfo{person}{Jun Xu}, \bibinfo{person}{Yue Yin}, \bibinfo{person}{Yang Song}, {and} \bibinfo{person}{Ji-Rong Wen}.} \bibinfo{year}{2022}\natexlab{}.
\newblock \showarticletitle{A Model-Agnostic Causal Learning Framework for Recommendation Using Search Data}. In \bibinfo{booktitle}{\emph{Proceedings of the ACM Web Conference 2022}} (Virtual Event, Lyon, France) \emph{(\bibinfo{series}{WWW '22})}. \bibinfo{publisher}{Association for Computing Machinery}, \bibinfo{address}{New York, NY, USA}, \bibinfo{pages}{224–233}.
\newblock
\showISBNx{9781450390965}


\bibitem[Si et~al\mbox{.}(2023)]%
        {SESRec}
\bibfield{author}{\bibinfo{person}{Zihua Si}, \bibinfo{person}{Zhongxiang Sun}, \bibinfo{person}{Xiao Zhang}, \bibinfo{person}{Jun Xu}, \bibinfo{person}{Xiaoxue Zang}, \bibinfo{person}{Yang Song}, \bibinfo{person}{Kun Gai}, {and} \bibinfo{person}{Ji-Rong Wen}.} \bibinfo{year}{2023}\natexlab{}.
\newblock \showarticletitle{When search meets recommendation: Learning disentangled search representation for recommendation}. In \bibinfo{booktitle}{\emph{Proceedings of the 46th International ACM SIGIR Conference on Research and Development in Information Retrieval}}. \bibinfo{pages}{1313--1323}.
\newblock


\bibitem[Sun et~al\mbox{.}(2019)]%
        {BERT4REC}
\bibfield{author}{\bibinfo{person}{Fei Sun}, \bibinfo{person}{Jun Liu}, \bibinfo{person}{Jian Wu}, \bibinfo{person}{Changhua Pei}, \bibinfo{person}{Xiao Lin}, \bibinfo{person}{Wenwu Ou}, {and} \bibinfo{person}{Peng Jiang}.} \bibinfo{year}{2019}\natexlab{}.
\newblock \showarticletitle{BERT4Rec: Sequential Recommendation with Bidirectional Encoder Representations from Transformer}. In \bibinfo{booktitle}{\emph{Proceedings of the 28th ACM International Conference on Information and Knowledge Management}} (Beijing, China) \emph{(\bibinfo{series}{CIKM '19})}. \bibinfo{publisher}{ACM}, \bibinfo{address}{New York, NY, USA}, \bibinfo{pages}{1441--1450}.
\newblock
\showISBNx{978-1-4503-6976-3}


\bibitem[Sun et~al\mbox{.}(2024)]%
        {sun2024learning}
\bibfield{author}{\bibinfo{person}{Weiwei Sun}, \bibinfo{person}{Lingyong Yan}, \bibinfo{person}{Zheng Chen}, \bibinfo{person}{Shuaiqiang Wang}, \bibinfo{person}{Haichao Zhu}, \bibinfo{person}{Pengjie Ren}, \bibinfo{person}{Zhumin Chen}, \bibinfo{person}{Dawei Yin}, \bibinfo{person}{Maarten Rijke}, {and} \bibinfo{person}{Zhaochun Ren}.} \bibinfo{year}{2024}\natexlab{}.
\newblock \showarticletitle{Learning to tokenize for generative retrieval}.
\newblock \bibinfo{journal}{\emph{Advances in Neural Information Processing Systems}}  \bibinfo{volume}{36} (\bibinfo{year}{2024}).
\newblock


\bibitem[Tang et~al\mbox{.}(2025)]%
        {tang2025thinkrecommendunleashinglatent}
\bibfield{author}{\bibinfo{person}{Jiakai Tang}, \bibinfo{person}{Sunhao Dai}, \bibinfo{person}{Teng Shi}, \bibinfo{person}{Jun Xu}, \bibinfo{person}{Xu Chen}, \bibinfo{person}{Wen Chen}, \bibinfo{person}{Wu Jian}, {and} \bibinfo{person}{Yuning Jiang}.} \bibinfo{year}{2025}\natexlab{}.
\newblock \bibinfo{title}{Think Before Recommend: Unleashing the Latent Reasoning Power for Sequential Recommendation}.
\newblock
\newblock
\showeprint[arxiv]{2503.22675}~[cs.IR]
\urldef\tempurl%
\url{https://arxiv.org/abs/2503.22675}
\showURL{%
\tempurl}


\bibitem[Tay et~al\mbox{.}(2022)]%
        {tay2022transformer}
\bibfield{author}{\bibinfo{person}{Yi Tay}, \bibinfo{person}{Vinh Tran}, \bibinfo{person}{Mostafa Dehghani}, \bibinfo{person}{Jianmo Ni}, \bibinfo{person}{Dara Bahri}, \bibinfo{person}{Harsh Mehta}, \bibinfo{person}{Zhen Qin}, \bibinfo{person}{Kai Hui}, \bibinfo{person}{Zhe Zhao}, \bibinfo{person}{Jai Gupta}, {et~al\mbox{.}}} \bibinfo{year}{2022}\natexlab{}.
\newblock \showarticletitle{Transformer memory as a differentiable search index}.
\newblock \bibinfo{journal}{\emph{Advances in Neural Information Processing Systems}}  \bibinfo{volume}{35} (\bibinfo{year}{2022}), \bibinfo{pages}{21831--21843}.
\newblock


\bibitem[Wang et~al\mbox{.}(2024b)]%
        {wang2024multilingual}
\bibfield{author}{\bibinfo{person}{Liang Wang}, \bibinfo{person}{Nan Yang}, \bibinfo{person}{Xiaolong Huang}, \bibinfo{person}{Linjun Yang}, \bibinfo{person}{Rangan Majumder}, {and} \bibinfo{person}{Furu Wei}.} \bibinfo{year}{2024}\natexlab{b}.
\newblock \showarticletitle{Multilingual e5 text embeddings: A technical report}.
\newblock \bibinfo{journal}{\emph{arXiv preprint arXiv:2402.05672}} (\bibinfo{year}{2024}).
\newblock


\bibitem[Wang et~al\mbox{.}(2024a)]%
        {wang2024enhancing}
\bibfield{author}{\bibinfo{person}{Yuening Wang}, \bibinfo{person}{Man Chen}, \bibinfo{person}{Yaochen Hu}, \bibinfo{person}{Wei Guo}, \bibinfo{person}{Yingxue Zhang}, \bibinfo{person}{Huifeng Guo}, \bibinfo{person}{Yong Liu}, {and} \bibinfo{person}{Mark Coates}.} \bibinfo{year}{2024}\natexlab{a}.
\newblock \showarticletitle{Enhancing Click-through Rate Prediction in Recommendation Domain with Search Query Representation}. In \bibinfo{booktitle}{\emph{Proceedings of the 33rd ACM International Conference on Information and Knowledge Management}}. \bibinfo{pages}{2462--2471}.
\newblock


\bibitem[Wang et~al\mbox{.}(2022)]%
        {wang2022neural}
\bibfield{author}{\bibinfo{person}{Yujing Wang}, \bibinfo{person}{Yingyan Hou}, \bibinfo{person}{Haonan Wang}, \bibinfo{person}{Ziming Miao}, \bibinfo{person}{Shibin Wu}, \bibinfo{person}{Qi Chen}, \bibinfo{person}{Yuqing Xia}, \bibinfo{person}{Chengmin Chi}, \bibinfo{person}{Guoshuai Zhao}, \bibinfo{person}{Zheng Liu}, {et~al\mbox{.}}} \bibinfo{year}{2022}\natexlab{}.
\newblock \showarticletitle{A neural corpus indexer for document retrieval}.
\newblock \bibinfo{journal}{\emph{Advances in Neural Information Processing Systems}}  \bibinfo{volume}{35} (\bibinfo{year}{2022}), \bibinfo{pages}{25600--25614}.
\newblock


\bibitem[Wang et~al\mbox{.}(2023)]%
        {wang2023exploiting}
\bibfield{author}{\bibinfo{person}{Yu Wang}, \bibinfo{person}{Zhengyang Wang}, \bibinfo{person}{Hengrui Zhang}, \bibinfo{person}{Qingyu Yin}, \bibinfo{person}{Xianfeng Tang}, \bibinfo{person}{Yinghan Wang}, \bibinfo{person}{Danqing Zhang}, \bibinfo{person}{Limeng Cui}, \bibinfo{person}{Monica Cheng}, \bibinfo{person}{Bing Yin}, {et~al\mbox{.}}} \bibinfo{year}{2023}\natexlab{}.
\newblock \showarticletitle{Exploiting intent evolution in e-commercial query recommendation}. In \bibinfo{booktitle}{\emph{Proceedings of the 29th ACM SIGKDD Conference on Knowledge Discovery and Data Mining}}. \bibinfo{pages}{5162--5173}.
\newblock


\bibitem[Xiao et~al\mbox{.}(2024)]%
        {xiao2024c}
\bibfield{author}{\bibinfo{person}{Shitao Xiao}, \bibinfo{person}{Zheng Liu}, \bibinfo{person}{Peitian Zhang}, \bibinfo{person}{Niklas Muennighoff}, \bibinfo{person}{Defu Lian}, {and} \bibinfo{person}{Jian-Yun Nie}.} \bibinfo{year}{2024}\natexlab{}.
\newblock \showarticletitle{C-pack: Packed resources for general chinese embeddings}. In \bibinfo{booktitle}{\emph{Proceedings of the 47th International ACM SIGIR Conference on Research and Development in Information Retrieval}}. \bibinfo{pages}{641--649}.
\newblock


\bibitem[Xie et~al\mbox{.}(2024)]%
        {xie2024unifiedssr}
\bibfield{author}{\bibinfo{person}{Jiayi Xie}, \bibinfo{person}{Shang Liu}, \bibinfo{person}{Gao Cong}, {and} \bibinfo{person}{Zhenzhong Chen}.} \bibinfo{year}{2024}\natexlab{}.
\newblock \showarticletitle{UnifiedSSR: A Unified Framework of Sequential Search and Recommendation}. In \bibinfo{booktitle}{\emph{Proceedings of the ACM on Web Conference 2024}}. \bibinfo{pages}{3410--3419}.
\newblock


\bibitem[Xiong et~al\mbox{.}(2020)]%
        {xiong2020approximate}
\bibfield{author}{\bibinfo{person}{Lee Xiong}, \bibinfo{person}{Chenyan Xiong}, \bibinfo{person}{Ye Li}, \bibinfo{person}{Kwok-Fung Tang}, \bibinfo{person}{Jialin Liu}, \bibinfo{person}{Paul Bennett}, \bibinfo{person}{Junaid Ahmed}, {and} \bibinfo{person}{Arnold Overwijk}.} \bibinfo{year}{2020}\natexlab{}.
\newblock \showarticletitle{Approximate nearest neighbor negative contrastive learning for dense text retrieval}.
\newblock \bibinfo{journal}{\emph{arXiv preprint arXiv:2007.00808}} (\bibinfo{year}{2020}).
\newblock


\bibitem[Yao et~al\mbox{.}(2021)]%
        {USER}
\bibfield{author}{\bibinfo{person}{Jing Yao}, \bibinfo{person}{Zhicheng Dou}, \bibinfo{person}{Ruobing Xie}, \bibinfo{person}{Yanxiong Lu}, \bibinfo{person}{Zhiping Wang}, {and} \bibinfo{person}{Ji-Rong Wen}.} \bibinfo{year}{2021}\natexlab{}.
\newblock \showarticletitle{USER: A Unified Information Search and Recommendation Model Based on Integrated Behavior Sequence}. In \bibinfo{booktitle}{\emph{Proceedings of the 30th ACM International Conference on Information ]\&amp; Knowledge Management}} (Virtual Event, Queensland, Australia) \emph{(\bibinfo{series}{CIKM '21})}. \bibinfo{publisher}{Association for Computing Machinery}, \bibinfo{address}{New York, NY, USA}, \bibinfo{pages}{2373–2382}.
\newblock
\showISBNx{9781450384469}


\bibitem[Yuan et~al\mbox{.}(2023)]%
        {yuan2023go}
\bibfield{author}{\bibinfo{person}{Zheng Yuan}, \bibinfo{person}{Fajie Yuan}, \bibinfo{person}{Yu Song}, \bibinfo{person}{Youhua Li}, \bibinfo{person}{Junchen Fu}, \bibinfo{person}{Fei Yang}, \bibinfo{person}{Yunzhu Pan}, {and} \bibinfo{person}{Yongxin Ni}.} \bibinfo{year}{2023}\natexlab{}.
\newblock \showarticletitle{Where to go next for recommender systems? id-vs. modality-based recommender models revisited}. In \bibinfo{booktitle}{\emph{Proceedings of the 46th International ACM SIGIR Conference on Research and Development in Information Retrieval}}. \bibinfo{pages}{2639--2649}.
\newblock


\bibitem[Yue et~al\mbox{.}(2024)]%
        {LRURec}
\bibfield{author}{\bibinfo{person}{Zhenrui Yue}, \bibinfo{person}{Yueqi Wang}, \bibinfo{person}{Zhankui He}, \bibinfo{person}{Huimin Zeng}, \bibinfo{person}{Julian McAuley}, {and} \bibinfo{person}{Dong Wang}.} \bibinfo{year}{2024}\natexlab{}.
\newblock \showarticletitle{Linear recurrent units for sequential recommendation}. In \bibinfo{booktitle}{\emph{Proceedings of the 17th ACM International Conference on Web Search and Data Mining}}. \bibinfo{pages}{930--938}.
\newblock


\bibitem[Zamani and Croft(2018)]%
        {JSR}
\bibfield{author}{\bibinfo{person}{Hamed Zamani} {and} \bibinfo{person}{W.~Bruce Croft}.} \bibinfo{year}{2018}\natexlab{}.
\newblock \showarticletitle{Joint Modeling and Optimization of Search and Recommendation}. In \bibinfo{booktitle}{\emph{Proceedings of the First Biennial Conference on Design of Experimental Search {\&} Information Retrieval Systems, Bertinoro, Italy, August 28-31, 2018}} \emph{(\bibinfo{series}{{CEUR} Workshop Proceedings}, Vol.~\bibinfo{volume}{2167})}. \bibinfo{publisher}{CEUR-WS.org}, \bibinfo{pages}{36--41}.
\newblock


\bibitem[Zamani and Croft(2020)]%
        {JSR2}
\bibfield{author}{\bibinfo{person}{Hamed Zamani} {and} \bibinfo{person}{W.~Bruce Croft}.} \bibinfo{year}{2020}\natexlab{}.
\newblock \showarticletitle{Learning a Joint Search and Recommendation Model from User-Item Interactions}. In \bibinfo{booktitle}{\emph{Proceedings of the 13th International Conference on Web Search and Data Mining}} (Houston, TX, USA) \emph{(\bibinfo{series}{WSDM '20})}. \bibinfo{publisher}{Association for Computing Machinery}, \bibinfo{address}{New York, NY, USA}, \bibinfo{pages}{717–725}.
\newblock
\showISBNx{9781450368223}


\bibitem[Zhang et~al\mbox{.}(2024c)]%
        {zhang2024modeling}
\bibfield{author}{\bibinfo{person}{Changshuo Zhang}, \bibinfo{person}{Teng Shi}, \bibinfo{person}{Xiao Zhang}, \bibinfo{person}{Qi Liu}, \bibinfo{person}{Ruobing Xie}, \bibinfo{person}{Jun Xu}, {and} \bibinfo{person}{Ji-Rong Wen}.} \bibinfo{year}{2024}\natexlab{c}.
\newblock \showarticletitle{Modeling Domain and Feedback Transitions for Cross-Domain Sequential Recommendation}.
\newblock \bibinfo{journal}{\emph{arXiv preprint arXiv:2408.08209}} (\bibinfo{year}{2024}).
\newblock


\bibitem[Zhang et~al\mbox{.}(2024d)]%
        {zhang2024qagcf}
\bibfield{author}{\bibinfo{person}{Changshuo Zhang}, \bibinfo{person}{Teng Shi}, \bibinfo{person}{Xiao Zhang}, \bibinfo{person}{Yanping Zheng}, \bibinfo{person}{Ruobing Xie}, \bibinfo{person}{Qi Liu}, \bibinfo{person}{Jun Xu}, {and} \bibinfo{person}{Ji-Rong Wen}.} \bibinfo{year}{2024}\natexlab{d}.
\newblock \showarticletitle{QAGCF: Graph Collaborative Filtering for Q\&A Recommendation}.
\newblock \bibinfo{journal}{\emph{arXiv preprint arXiv:2406.04828}} (\bibinfo{year}{2024}).
\newblock


\bibitem[Zhang et~al\mbox{.}(2025)]%
        {zhang2025testtimealignmenttrackinguser}
\bibfield{author}{\bibinfo{person}{Changshuo Zhang}, \bibinfo{person}{Xiao Zhang}, \bibinfo{person}{Teng Shi}, \bibinfo{person}{Jun Xu}, {and} \bibinfo{person}{Ji-Rong Wen}.} \bibinfo{year}{2025}\natexlab{}.
\newblock \bibinfo{title}{Test-Time Alignment for Tracking User Interest Shifts in Sequential Recommendation}.
\newblock
\newblock
\showeprint[arxiv]{2504.01489}~[cs.IR]
\urldef\tempurl%
\url{https://arxiv.org/abs/2504.01489}
\showURL{%
\tempurl}


\bibitem[Zhang et~al\mbox{.}(2024a)]%
        {zhang2024saqrec}
\bibfield{author}{\bibinfo{person}{Kepu Zhang}, \bibinfo{person}{Teng Shi}, \bibinfo{person}{Sunhao Dai}, \bibinfo{person}{Xiao Zhang}, \bibinfo{person}{Yinfeng Li}, \bibinfo{person}{Jing Lu}, \bibinfo{person}{Xiaoxue Zang}, \bibinfo{person}{Yang Song}, {and} \bibinfo{person}{Jun Xu}.} \bibinfo{year}{2024}\natexlab{a}.
\newblock \showarticletitle{SAQRec: Aligning Recommender Systems to User Satisfaction via Questionnaire Feedback}. In \bibinfo{booktitle}{\emph{Proceedings of the 33rd ACM International Conference on Information and Knowledge Management}}. \bibinfo{pages}{3165--3175}.
\newblock


\bibitem[Zhang et~al\mbox{.}(2024b)]%
        {zhang2024model}
\bibfield{author}{\bibinfo{person}{Xiao Zhang}, \bibinfo{person}{Teng Shi}, \bibinfo{person}{Jun Xu}, \bibinfo{person}{Zhenhua Dong}, {and} \bibinfo{person}{Ji-Rong Wen}.} \bibinfo{year}{2024}\natexlab{b}.
\newblock \showarticletitle{Model-Agnostic Causal Embedding Learning for Counterfactually Group-Fair Recommendation}.
\newblock \bibinfo{journal}{\emph{IEEE Transactions on Knowledge and Data Engineering}} (\bibinfo{year}{2024}).
\newblock


\bibitem[Zhang et~al\mbox{.}(2024e)]%
        {zhang2024unified}
\bibfield{author}{\bibinfo{person}{Yuting Zhang}, \bibinfo{person}{Yiqing Wu}, \bibinfo{person}{Ruidong Han}, \bibinfo{person}{Ying Sun}, \bibinfo{person}{Yongchun Zhu}, \bibinfo{person}{Xiang Li}, \bibinfo{person}{Wei Lin}, \bibinfo{person}{Fuzhen Zhuang}, \bibinfo{person}{Zhulin An}, {and} \bibinfo{person}{Yongjun Xu}.} \bibinfo{year}{2024}\natexlab{e}.
\newblock \showarticletitle{Unified Dual-Intent Translation for Joint Modeling of Search and Recommendation}. In \bibinfo{booktitle}{\emph{Proceedings of the 30th ACM SIGKDD Conference on Knowledge Discovery and Data Mining}}. \bibinfo{pages}{6291--6300}.
\newblock


\bibitem[Zhao et~al\mbox{.}(2022)]%
        {SRJgraph}
\bibfield{author}{\bibinfo{person}{Kai Zhao}, \bibinfo{person}{Yukun Zheng}, \bibinfo{person}{Tao Zhuang}, \bibinfo{person}{Xiang Li}, {and} \bibinfo{person}{Xiaoyi Zeng}.} \bibinfo{year}{2022}\natexlab{}.
\newblock \showarticletitle{Joint Learning of E-Commerce Search and Recommendation with a Unified Graph Neural Network}. In \bibinfo{booktitle}{\emph{Proceedings of the Fifteenth ACM International Conference on Web Search and Data Mining}} (Virtual Event, AZ, USA) \emph{(\bibinfo{series}{WSDM '22})}. \bibinfo{publisher}{Association for Computing Machinery}, \bibinfo{address}{New York, NY, USA}, \bibinfo{pages}{1461–1469}.
\newblock
\showISBNx{9781450391320}


\bibitem[Zhao et~al\mbox{.}(2023)]%
        {zhao2023survey}
\bibfield{author}{\bibinfo{person}{Wayne~Xin Zhao}, \bibinfo{person}{Kun Zhou}, \bibinfo{person}{Junyi Li}, \bibinfo{person}{Tianyi Tang}, \bibinfo{person}{Xiaolei Wang}, \bibinfo{person}{Yupeng Hou}, \bibinfo{person}{Yingqian Min}, \bibinfo{person}{Beichen Zhang}, \bibinfo{person}{Junjie Zhang}, \bibinfo{person}{Zican Dong}, {et~al\mbox{.}}} \bibinfo{year}{2023}\natexlab{}.
\newblock \showarticletitle{A survey of large language models}.
\newblock \bibinfo{journal}{\emph{arXiv preprint arXiv:2303.18223}} (\bibinfo{year}{2023}).
\newblock


\bibitem[Zheng et~al\mbox{.}(2024)]%
        {LC_Rec}
\bibfield{author}{\bibinfo{person}{Bowen Zheng}, \bibinfo{person}{Yupeng Hou}, \bibinfo{person}{Hongyu Lu}, \bibinfo{person}{Yu Chen}, \bibinfo{person}{Wayne~Xin Zhao}, \bibinfo{person}{Ming Chen}, {and} \bibinfo{person}{Ji-Rong Wen}.} \bibinfo{year}{2024}\natexlab{}.
\newblock \showarticletitle{Adapting large language models by integrating collaborative semantics for recommendation}. In \bibinfo{booktitle}{\emph{2024 IEEE 40th International Conference on Data Engineering (ICDE)}}. IEEE, \bibinfo{pages}{1435--1448}.
\newblock


\bibitem[Zhou et~al\mbox{.}(2022)]%
        {FMLPREC}
\bibfield{author}{\bibinfo{person}{Kun Zhou}, \bibinfo{person}{Hui Yu}, \bibinfo{person}{Wayne~Xin Zhao}, {and} \bibinfo{person}{Ji-Rong Wen}.} \bibinfo{year}{2022}\natexlab{}.
\newblock \showarticletitle{Filter-Enhanced MLP is All You Need for Sequential Recommendation}. In \bibinfo{booktitle}{\emph{Proceedings of the ACM Web Conference 2022}} (Virtual Event, Lyon, France) \emph{(\bibinfo{series}{WWW '22})}. \bibinfo{publisher}{Association for Computing Machinery}, \bibinfo{address}{New York, NY, USA}, \bibinfo{pages}{2388–2399}.
\newblock
\showISBNx{9781450390965}


\bibitem[Zhou et~al\mbox{.}(2023)]%
        {zhou2023webultron}
\bibfield{author}{\bibinfo{person}{Yujia Zhou}, \bibinfo{person}{Jing Yao}, \bibinfo{person}{Ledell Wu}, \bibinfo{person}{Zhicheng Dou}, {and} \bibinfo{person}{Ji-Rong Wen}.} \bibinfo{year}{2023}\natexlab{}.
\newblock \showarticletitle{WebUltron: An Ultimate Retriever on Webpages Under the Model-Centric Paradigm}.
\newblock \bibinfo{journal}{\emph{IEEE Transactions on Knowledge and Data Engineering}} (\bibinfo{year}{2023}).
\newblock


\bibitem[Zhuang et~al\mbox{.}(2022)]%
        {zhuang2022bridging}
\bibfield{author}{\bibinfo{person}{Shengyao Zhuang}, \bibinfo{person}{Houxing Ren}, \bibinfo{person}{Linjun Shou}, \bibinfo{person}{Jian Pei}, \bibinfo{person}{Ming Gong}, \bibinfo{person}{Guido Zuccon}, {and} \bibinfo{person}{Daxin Jiang}.} \bibinfo{year}{2022}\natexlab{}.
\newblock \showarticletitle{Bridging the gap between indexing and retrieval for differentiable search index with query generation}.
\newblock \bibinfo{journal}{\emph{arXiv preprint arXiv:2206.10128}} (\bibinfo{year}{2022}).
\newblock


\end{thebibliography}

\end{document}